\documentclass[sigconf]{acmart}

\newcommand{\system}{\textsc{Cocoa}\xspace}
\newcommand{\name}{Nora\xspace}

\AtBeginDocument{%
  \providecommand\BibTeX{{%
    \normalfont B\kern-0.5em{\scshape i\kern-0.25em b}\kern-0.8em\TeX}}}

\copyrightyear{2026}
\acmYear{2026}
\setcopyright{cc}
\setcctype{by}
\acmConference[CHI '26]{Proceedings of the 2026 CHI Conference on Human Factors in Computing Systems}{April 13--17, 2026}{Barcelona, Spain}
\acmBooktitle{Proceedings of the 2026 CHI Conference on Human Factors in Computing Systems (CHI '26), April 13--17, 2026, Barcelona, Spain}
\acmDOI{10.1145/3772318.3791673}
\acmISBN{979-8-4007-2278-3/2026/04}

\usepackage{enumitem}
\usepackage{graphicx}
\usepackage{url}
\usepackage{booktabs}  
\usepackage{caption}
\usepackage{subcaption}
\usepackage{xspace}
\usepackage{listings}
\usepackage{xcolor}

\graphicspath{{figures/}{pictures/}{images/}{./}}
\begin{document}

\title{\system: \underline{Co}-Planning and \underline{Co}-Execution with AI \underline{A}gents}

\author{K. J. Kevin Feng}
\authornote{Work completed during an internship at Ai2.}
\affiliation{%
  \institution{University of Washington}
  \city{Seattle, WA}
  \country{USA}
  }
\email{kjfeng@uw.edu}

\author{Kevin Pu}
\authornotemark[1]
\affiliation{%
  \institution{University of Toronto}
  \city{Toronto, ON}
  \country{Canada}
  }
\email{jpu@dgp.toronto.edu}

\author{Matt Latzke}
\affiliation{%
  \institution{Allen Institute for AI}
  \city{Seattle, WA}
  \country{USA}
  }
\email{mattl@allenai.org}

\author{Tal August}
\affiliation{%
  \institution{UIUC}
  \city{Urbana, IL}
  \country{USA}
  }
\email{taugust@illinois.edu}

\author{Pao Siangliulue}
\affiliation{%
  \institution{Allen Institute for AI}
  \city{Seattle, WA}
  \country{USA}
  }
\email{paos@allenai.org}

\author{Jonathan Bragg}
\affiliation{%
  \institution{Allen Institute for AI}
  \city{Seattle, WA}
  \country{USA}
  }
\email{jbragg@allenai.org}

\author{Daniel S. Weld}
\affiliation{%
  \institution{Allen Institute for AI}
  \city{Seattle, WA}
  \country{USA}
  }
\email{danw@allenai.org}

\author{Amy X. Zhang}
\affiliation{%
  \institution{University of Washington}
  \city{Seattle, WA}
  \country{USA}
  }
\email{axz@cs.uw.edu}

\author{Joseph Chee Chang}
\affiliation{%
  \institution{Allen Institute for AI}
  \city{Seattle, WA}
  \country{USA}
  }
\email{josephc@allenai.org}

\begin{teaserfigure}
  \centering
  \includegraphics[width=1\textwidth]{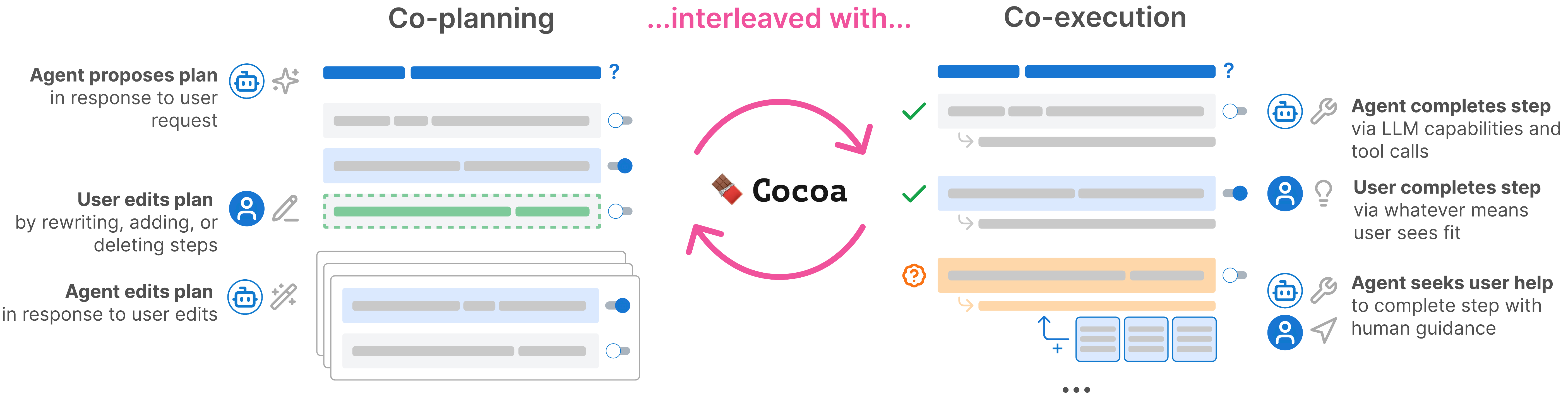}
  \caption{\system is an interactive system that facilitates interleaved \underline{co}-planning and \underline{co}-execution with AI \underline{a}gents in a document environment for scientific researchers. \system integrates AI agents into documents through an interface where a human user and an AI agent can jointly plan and execute plan steps using a shared representation of tasks, roles, and progress directly in the document.}
  \label{fig:teaser-abstract}
  \Description{On the left: an illustration of co-planning: the agent proposes a plan of action in response to a user request. The user then edits the plan by rewriting, adding, or deleting steps, and the agent can further edit the plan by responding to user requests. On the right: an illustration of co-execution. an agent complete steps it is assigned to, and the user completes steps assigned to them. The user can also help the agent complete its assigned steps if the agent fails to do so. Cocoa is a system that interleaves co-planning and co-execution.}
\end{teaserfigure}

\renewcommand{\shortauthors}{Feng, et al.}

\begin{abstract}

As AI agents take on increasingly long-running tasks involving sophisticated planning and execution, there is a corresponding need for novel interaction designs that enable deeper human-agent \emph{collaboration}. However, most prior works leverage human interaction to fix ``autonomous'' workflows that have yet to become fully autonomous or rigidly treat planning and execution as separate stages. Based on a formative study with 9 researchers using AI to support their work, we propose a design that affords greater flexibility in collaboration, so that users can 1) \textit{delegate agency} to the user or agent via a collaborative plan where individual steps can be assigned; and
2) \emph{interleave} planning and execution so that plans can adjust after partial execution.
We introduce \textsc{Cocoa}, a system that takes design inspiration from computational notebooks to support complex research tasks. A lab study ($n=16$) found that \textsc{Cocoa} enabled steerability without sacrificing ease-of-use, and a week-long field deployment ($n=7$) showed how researchers collaborated with \textsc{Cocoa} to accomplish real-world tasks. 


\end{abstract}

\maketitle

\section{Introduction}

\begin{figure*}[h!]
    \includegraphics[width=1\linewidth]{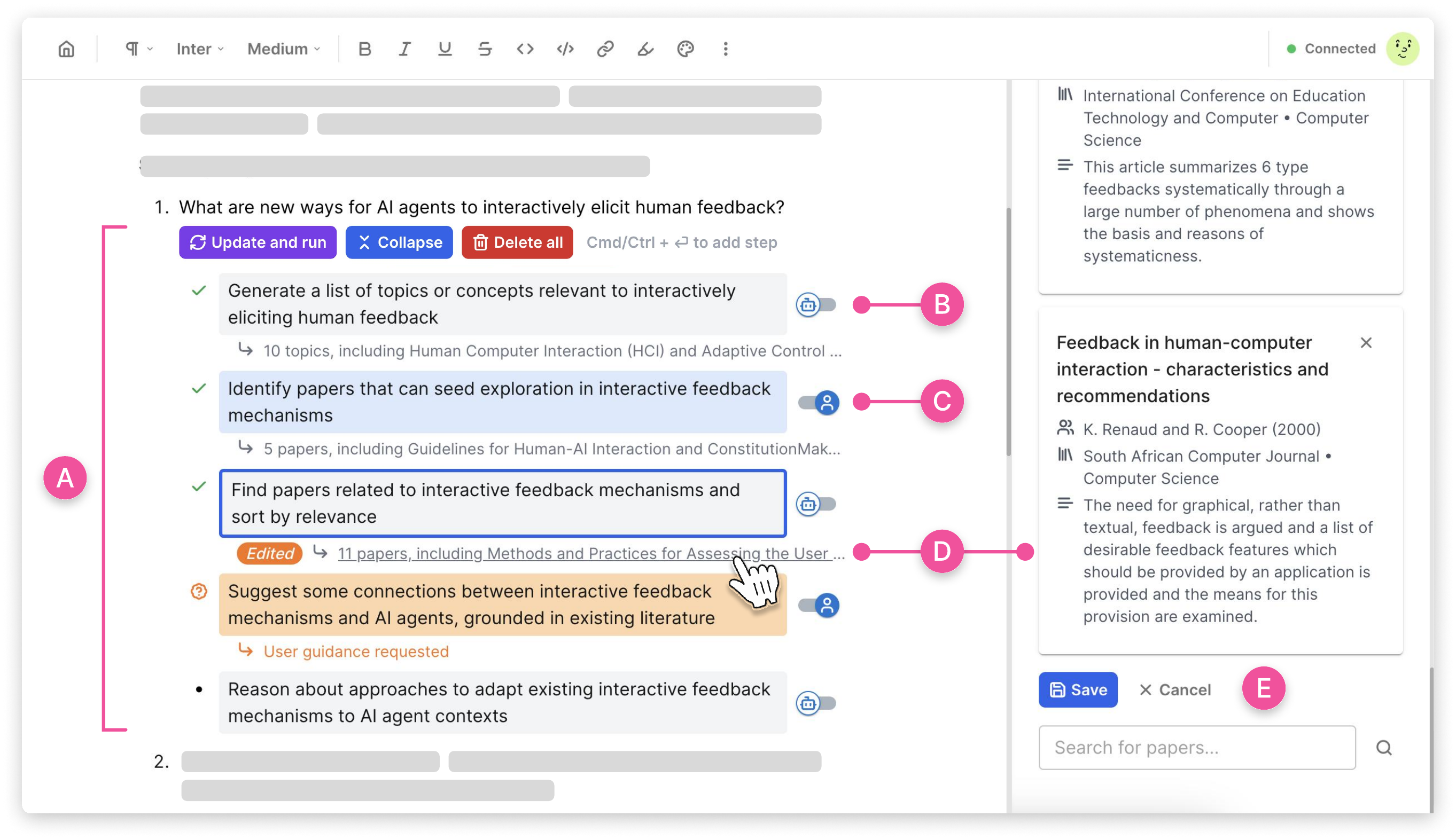}
    \caption{An overview of the \system user interface. An interactive plan (A) affords human-agent co-planning and co-execution: a researcher and the AI agent can collaboratively edit the plan in the document and execute the plan steps, similar to executing code cells in a computational notebook. Steps can be assigned to the AI agent (B) or the researcher (C). The researcher can freely edit the AI agent's outputs in an interactive sidebar (D), such as adding relevant papers that the agent did not find (E) to help steer the agent with their feedback and expertise. In this example, the first three steps of a plan to summarize methods for human feedback elicitation have already been executed, and the agent is requesting guidance from the user in the next step.}
    \label{fig:teaser-detailed}
    \Description{An overview of the Cocoa user interface, which resembles a document editor. An interactive plan affords human-agent co-planning and co-execution: a researcher and the AI agent can collaboratively edit the plan in the document and execute the plan steps, similar to executing code cells in a computational notebook. Steps can be assigned to the AI agent or the researcher. The researcher can freely edit the AI agent's outputs in an interactive sidebar, such as adding relevant papers that the agent did not find, to help steer the agent with their feedback and expertise. In the example shown, the first three steps of a plan to summarize methods for human feedback elicitation have already been executed, and the agent is requesting guidance from the user in the next step.}
\end{figure*}

Since the advent of personal computing, researchers and practitioners in artificial intelligence (AI) and human-computer interaction (HCI) have set sights on developing AI agents that can collaborate with humans to perform complex computer-based tasks \cite{Maes1994AgentsTR, Maes1996InterfaceA, Lieberman1997AutonomousIA, Horvitz1999PrinciplesOM}. For example, paradigms such as mixed-initiative interaction envisioned fluid human-AI collaboration with user-led direct data manipulation combined with AI-led autonomous task execution \cite{Horvitz1999PrinciplesOM}. 
Recent advancements in large language models (LLMs) and agent frameworks involving reasoning, planning, memory, and tool use \cite{Yao2022ReActSR, Sumers2023CognitiveAF, Schick2023ToolformerLM, Wang2023VoyagerAO, Kim2024HuskyAU} have accelerated the development of fully autonomous AI agents to tackle complex tasks, including shopping online \cite{yao2022webshop}, writing software \cite{Yang2024SWEagentAI, Jimenez2023SWEbenchCL, devin-ai}, performing ``deep research'' to assemble a report based on web sources \cite{gemini-deep-research, openai-deep-research, perplexity-deep-research}, and general computer use \cite{anthropic-computer-use, openai-operator, shaikh2025creating}. 
Despite the push towards full autonomy, recent work has recognized the importance of deeper human-agent \textit{collaboration} for not only bolstering agent performance on complex tasks \cite{Shi2024CanLM}, but also improving accessibility \cite{Peng2025MoraePP}, user control and trust \cite{Mozannar2025MagenticUITH, Huq2025CowPilotAF}, safety \cite{openai-operator}, and user agency more generally \cite{feng2025levels}, 
leading to the development of more collaborative human-agent systems, such as for web browsing \cite{Peng2025MoraePP, Huq2025CowPilotAF, Mozannar2025MagenticUITH} and general productivity \cite{Shao2024CollaborativeGA, openai-operator}. 

However, many of these systems incorporate prescribed collaborative workflows that may limit their utility and steerability.
Some works use \textit{agent-guided workflows}: the agent dictates what input to elicit from the user, as well as when and how the elicitation happens~\cite{Peng2025MoraePP, Huq2025CowPilotAF, openai-operator, Mozannar2025MagenticUITH, Shao2024CollaborativeGA}. Users may intervene by manually interrupting or taking control of the agent \cite{Mozannar2025MagenticUITH, openai-operator, openai-deep-research}, but these efforts are reactive and rely on persistent awareness of where the agent has erred.  
On the other hand, researchers have developed systems with \textit{user-guided workflows}, where the user is in full control of task planning, while the agent executes scoped subtasks \cite{he2025plan, Kazemitabaar2024ImprovingSA, Suh2023SensecapeEM, Angert2023SpellburstAN}. Whether agent- or user-guided, these systems encode a \textit{particular configuration} of human-agent collaboration---there are few affordances to 
\textit{flexibly delegate agency} by altering who is guiding based on environmental context or user preferences~\cite{satyanarayan2024intelligence}.

Separately, researchers and industry practitioners have realized the value of decoupling human-agent planning and execution to improve control and avoid fruitless execution \cite{chip-agents, he2025plan, Kazemitabaar2024ImprovingSA}. However, completely isolating the two into rigid stages can lead to poor results as well---plans often need to be iteratively adapted based on results from completed steps \cite{Mozannar2025MagenticUITH}. In contrast, fully autonomous frameworks like ReAct \cite{Yao2022ReActSR} \textit{interleaves} planning and execution, and has shown benefits in allowing fully autonomous AI agents to readjust their plans after seeing partial execution results. We thus see a new design opportunity for human-agent interaction and ask: \textbf{what might flexible configurations of agency in human-agent collaboration across planning and execution look like, and how will users make use of them in their own work?}

In this work, we explore a novel instantiation of mixed-initiative interaction \cite{Horvitz1999PrinciplesOM} for \textit{explicitly and flexibly delegating work} in human-agent hybrid workflows by \textit{interleaving collaborative planning and execution.} We conduct this exploration in the domain of scientific research, a challenging domain for human-agent collaboration due to the prevalence of knowledge-intensive workflows and nuanced decisions made with domain-specific knowledge. Existing agentic research tools primarily focus on supporting human-agent collaboration either \textit{before} or \textit{after} task execution. One group of systems elicits user intent and preferences before execution \cite{openai-deep-research, gemini-deep-research, perplexity-deep-research} (e.g., asking clarification questions about users' goals \cite{openai-deep-research}, allowing plan review and modification \cite{gemini-deep-research, Mozannar2025MagenticUITH}). Conversely, another group helps users verify and correct the agent after it has generated an output after executing its plan \cite{openai-canvas, guo2025deepseek}. In practice, because planning and execution are deeply interleaved, we aim to support human-agent collaboration throughout \textit{fluid movements} between the two phases.

We introduce \system, an interactive system that implements our approach by supporting \emph{hybrid human-agent workflows} for scientific researchers in a document editor---a common site for research work. This interleaving happens via the following three ways. First, \system supports human-agent \textbf{co-planning}: the agent, when invoked in the document, will first propose a plan of action that is fully interactive and integrated within the document. The user can then edit steps, add new ones, and assign each step to either themselves or the agent. Second, \system supports \textbf{human-agent co-execution}: drawing inspiration from the design of computational notebooks, \system allows the user and the agent to collaboratively complete plan steps, re-executing steps as desired. The user can interactively refine the agent’s intermediate outputs and also manually take over steps themselves to steer the workflow in a more desirable direction. Third, and most importantly, \system \textbf{interleaves co-planning and co-execution}: users can smoothly transition between the two and modify their plans based on outputs from execution, and vice versa.

We then evaluate \system through a lab study with 16 researchers, where we found that \textit{interleaving co-planning and co-execution} in \system afforded greater agent steerability while maintaining ease of use compared to a strong chat baseline. We finally conducted a 7-day field deployment of \system with 7 researchers, where they found the \textit{explicit delegation of work between themselves and the agent}, in addition to interleaved co-planning and co-execution, to be valuable in their day-to-day research.

Concretely, our work makes the following contributions:

\begin{enumerate}
    \item A formative study with 9 researchers that motivates the design of interactive systems that can flexibly delegate work between the user and agent across task planning and execution. 
    \item \system, an interactive system that interleaves human-agent co-planning and co-execution in a document editor for scientific researchers.
     \item A lab evaluation of \system with 16 researchers, where we found that  participants frequently made use of interleaving co-planning and co-execution when completing research tasks, suggesting that the two are mutually beneficial while enabling more granular steering of the agent.
     \item A 7-day deployment study of \system with 7 researchers, where participants made use of self- and AI-assigned steps to flexibly delegate human-AI agency and embed their expertise into the  workflow.  
     
\end{enumerate}

\section{Related Work}

\subsection{Planning and Interactivity in LLM Agents}
 
Prior work has identified core components in LLM agent architectures to consist of memory, reasoning, planning, and tool use \cite{Yao2022ReActSR, Zhou2023LanguageAT, liu2024agent, Sumers2023CognitiveAF, Wang2023VoyagerAO, Kim2024HuskyAU}.  
Central to LLM agents' longitudinal operations is long-term planning, which demands capable reasoning abilities \cite{Hao2023ReasoningWL} and thus is often implemented by chain-of-thought (CoT)\footnote{We use CoT as an umbrella term encompassing chain-of-thought, trees-of-thought \cite{Yao2023TreeOT, Zhuang2023ToolChainEA}, and related methods.} \cite{Wei2022ChainOT, Qin2023ToolLLMFL, Yao2022ReActSR, openai-o1}. CoT provides a series of intermediate reasoning steps as exemplars in prompting to boost performance on complex reasoning tasks \cite{Wei2022ChainOT, openai-o1}. CoT is crucial for facilitating task decomposition and therefore LLM agents' planning capabilities \cite{Qin2023ToolLLMFL, chen2023put}. However, even with CoT, critical investigations of LLMs' abilities to autonomously generate executable plans reveal limited success across diverse domains \cite{Valmeekam2023OnTP}. Indeed, although CoT can help improve model planning for tasks with well-defined, objective solutions that provide clear signals for reinforcement learning---e.g., numerical, tabular, and knowledge-based reasoning \cite{Kim2024HuskyAU}---it may not be so effective in domains that require expert tacit knowledge to navigate ambiguous problem spaces with no single right answer \cite{Hao2023ReasoningWL}. Unlike model properties that show empirical improvement through scaling laws, limitations of planning in these domains may not resolve with scale alone as 1) tacit knowledge is not well-documented in training data and is thus difficult for a model to robustly learn \cite{Cheong2024AIAN, Feng2023CaseRT, Zhang2024MASSWAN}, and 2) there may be no ``correct'' workflow for CoT to follow and verify the correctness post-hoc \cite{jirotka2013supporting}. Scientific research is one such domain \cite{Zhang2024MASSWAN}.

In light of this, recent work has recognized the need to interactively incorporate user feedback for improving LLM agents' planning capabilities and beyond \cite{Kazemitabaar2024ImprovingSA, Lawley2023VALIT, Shao2024CollaborativeGA, openai-operator, Huq2025CowPilotAF, Mozannar2025MagenticUITH, Peng2025MoraePP}, particularly within scientific discovery \cite{Majumder2024DatadrivenDW, gottweis2025towards,Jansen2025CodeScientistES}. For example,
Lawley and MacLellan \cite{Lawley2023VALIT} architected an approach where user interaction is used to guide the model in planning for unseen tasks on-the-fly using a hierarchical network of smaller actions. OpenAI's Operator \cite{openai-operator} allows users to ``take control'' of the agent to perform tasks manually. Yet, interactive techniques in this area are still nascent, despite the new challenges identified for human-agent communication \cite{bansal2024challenges}. Interfaces for LLM assistants (e.g., AgentGPT \cite{agentgpt}, Devin \cite{devin-ai}) have been primarily limited to chat interfaces targeted at \textit{monitoring} agent activity rather than \textit{empowering the user to proactively collaborate} with the agent. 

An emerging body of work has started to experiment with more interactive techniques for agents. He et al. \cite{he2025plan} separated agent planning from execution and found that user involvement during execution can significantly boost agent performance, but poses a high cognitive load. Kazemitabaar et al. \cite{Kazemitabaar2024ImprovingSA} developed interfaces for data analysts to edit an execution plan of an LLM to provide more control points for steering behavior, while Google's Deep Research \cite{gemini-deep-research} can re-generate a plan if the user provides feedback via prompting. However, these interactions are \textit{corrective} rather than \textit{collaborative}---that is, users would typically engage in these interactions to correct agent behavior post-hoc rather than \textit{proactively iterate back-and-forth} with the agent at multiple points in the workflow. We do this by introducing affordances for collaborative human-AI planning and execution via a shared representation of tasks and delegation of agency. While our work shares some conceptual similarities with Kazemitabaar et al. \cite{Kazemitabaar2024ImprovingSA}, we focus on enabling human-AI co-agency \textit{across steps} in multi-step plans rather than revealing hidden model reasoning \textit{at each step}.

\subsection{Computational Notebooks}
Donald Knuth's vision of \textit{literate programming} interleaves computer programs with natural language documentation \cite{Knuth1984LiterateP}. This vision later gave rise to the computational notebook paradigm, which organizes program imperatives, input data, and rich outputs (e.g., tables, visualizations, interactive widgets) into linearly arranged cells \cite{Chattopadhyay2020WhatsWW, lau2020notebooks}.
Computational notebooks are a well-studied paradigm in HCI (e.g., \cite{Choi2023TowardsTR, Ayobi2023ComputationalNA, Zheng2022TellingSF, Lin2023InkSightLS, Chattopadhyay2020WhatsWW, TranOLeary2023ImprimerCN, li2023notable}), with most prior work focusing on  how data scientists leverage them in their workflows.

In our work, we take inspiration from design decisions made for computational notebooks because of the many parallels between workflows in notebooks and mixed-initiative task completion systems \cite{Horvitz1999PrinciplesOM, GrundeMcLaughlin2023DesigningLC, Wu2023LLMsAW}. For instance, data scientists often break down a high-level data analysis task into executable cells in computational notebooks \cite{Kery2018TheSI, Rule2018ExplorationAE}.
As they execute each cell and inspect outputs, they can verify the quality and sensibility of outputs before moving forward in their analysis, or go back to modify existing code cells before execution to interleave between programming, execution, and output examination \cite{Kery2020mageFM}. Finally, the notebook can serve as documentation and communication of their activities \cite{li2023notable}.

By drawing these parallels, we can reveal new opportunities to design interactive workflows and interfaces for LLM agents. Just like how a data scientist may easily add new cells in a notebook or reconfigure existing cells to adapt their analysis plan, a user may interactively edit an agent's plan of execution to better steer the agent in productive directions. Furthermore, an agentic system may simultaneously be more usable and resource-efficient if a user could direct an agent to iterate on a subtask to improve its output before continuing onto subtasks dependent on that output. This also demands new ways of viewing and editing agent outputs on subtasks, which increases opportunities for human input over existing approaches such as simply logging agent actions \cite{autogpt, agentgpt, oai-assistants}. Our work exploits these parallels to contribute new design patterns for steering agentic systems. We also apply novel extensions of these parallels for human-agent collaboration: agent-proposed plans (i.e, dynamic AI-generated notebook cells) and agent vs. user steps (i.e, AI executes some cells while the user executes others). 

\subsection{AI for Scientific Research}
Significant efforts in recent years have advanced how AI can be used to support scientific research. 
Of particular interest to us are works that leverage these recent advances to help researchers with planning and execution of \textit{literature-focused tasks}. Planning a research project is a cognitively demanding, complex process consisting of iterative cycles of divergent and convergent thinking grounded in literature review \cite{Palani2023RelatedlySL, Dean2006IdentifyingQN, Chang2023CiteSeeAC}, making it an ideal scenario to explore human-agent collaboration.
One major thread of prior research has focused on leveraging AI models as tools or components in user-driven systems. These include tools for paper reading \cite{Chang2023CiteSeeAC, Rachatasumrit2022CiteReadIL} and skimming \cite{Fok2022ScimIS}, literature review \cite{Lee2024PaperWeaverET, Palani2023RelatedlySL}, paper recommendation \cite{Kang2022FromWY, Kang2022ThreddyAI}, information retrieval and sensemaking \cite{chan2018solvent, Kang2023SynergiAM, Fok2023QlarifyRE, Wang2024SciDaSynthIS}, and ideation \cite{Gu2024GenerationAH, Baek2024ResearchAgentIR, jansen2024discoveryworld}.
On the other end of the spectrum are fully automated end-to-end systems where an AI agent has much more agency in planning and executing literature review tasks \cite{scite, elicit}, or even to attempt carrying out entire research projects on their own \cite{lu2024ai, Jansen2025CodeScientistES, gottweis2025towards,gemini-deep-research, openai-deep-research, perplexity-deep-research, schmidgall2025agent}.

As AI systems become more capable, we ask: Is the generation of full research artifacts (e.g., research questions, proposals) the most desired and promising paradigm to assist researchers? Indeed, when working with other (human) collaborators, researchers often find richness in co-evolving partially completed artifacts \cite{star1989institutional, lee2007boundary, shrum2007structures, jirotka2013supporting} and answering or asking questions that stimulate critical thinking and reflection \cite{overholser1993elements, padesky1993socratic}. Further, some fully automatically generated artifacts are still found to be of lower quality than those generated with human involvement \cite{Jansen2025CodeScientistES}. Schmidgall et al. found that even systems designed for end-to-end research automation produce significantly higher quality outputs with human involvement at each stage \cite{schmidgall2025agent}.

In this work, we build on recent advances in AI research support tools, but rather than focusing on generating predetermined research artifacts, we contribute an interaction design pattern that allows for flexible specification of the final artifact and collaboration between a researcher and an AI agent.

\section{Formative Study}
\label{s:formative}

Building on prior work, we aim to enhance collaboration between humans and AI agents using a shared, interactive operational representation. We focus on scientific research---a challenging domain for human-agent collaboration given its complex, multi-step workflows and the nuanced decisions researchers make upon processing information-dense scientific papers. Specifically, we investigate the potential for project documents---continuous records of ideas, updates, and tasks---to act as the agent environment. We thus conducted a formative study to answer the following research questions:

\begin{itemize}
    \item[\textbf{F1:}] What are the properties and opportunities of researchers' project documents for human-agent collaboration? 
    \item[\textbf{F2:}] How do researchers engage in planning within project documents? 
    \item[\textbf{F3:}] How would researchers prefer to collaborate with an AI agent in their research workflows?
\end{itemize}

\subsection{Participants, Procedure, and Data Analysis}

We recruited 9 Ph.D. students (detailed demographics in Table \ref{t:participants-formative} of Appendix \ref{a:formative-study-participants}) through an interest form sent to a research organization's internal Slack channel and the authors' professional connections. We targeted Ph.D. students as they often lead the detailed planning and execution of research projects and are the primary editors of project documents.

All studies were conducted virtually over Google Meet and lasted 60 minutes. Before the study, we collected from participants an active or past research project document and a brief description of their research interests. The study was divided into 3 parts: 
\begin{enumerate}
    \item An activity involving the participant's project document to better understand their current planning behavior (25 minutes).
    \item An activity with a Wizard-of-Oz (WoZ) design probe to explore how researchers plan and explore research ideas in a document editor with LLM support via a chatbot (25 minutes).
    \item An exit interview where researchers reflected on their experience with the probe and using LLMs in research (10 minutes).
\end{enumerate}

Details for each part can be found in Appendix \ref{a:formative-procedure-details}. Each participant was given a \$35 USD honorarium after the study. The study was reviewed and exempted by our organization's internal review board. We qualitatively analyzed project documents produced from the study probe activity alongside study transcripts. Our data analysis methods can be found in Appendix \ref{a:formative-data-analysis}.

\subsection{Findings}
In this section, we redact any project-specific details for privacy and intellectual confidentiality.

\subsubsection{\textbf{[F1 \& F3] Project documents are promising environments for human-agent collaboration}}
\label{s:formative-documents}

Participants' project documents acted as a ``hub'' for their projects. These documents commonly included meeting notes, \emph{planned to-do items}, \emph{progress updates} tied to to-do items, \emph{research questions} to discuss with collaborators, and \emph{links to other documents} detailing a particular project component in greater depth. These documents often served as ``scratch paper'' for researchers to informally leave a trace of their reasoning and high-level goals, conduct short-term and long-term planning, and progress tracking of these plans. Most documents (P1, P2--4, P7--P9) contained one or more problem statements that motivated the project and stated its core contributions. From there, participants listed subgoals and ideas for planned exploration (all participants). 

Participants expressed a desire for tighter integration between an AI agent and their project documents in their research workflows for two main reasons. \textbf{First}, they wanted to receive in-situ AI support as they plan and execute research tasks in their documents. P1 and P2 both envisioned an AI system \textit{``actively engaging with the content I'm writing [...] after I write each statement, the system could retrieve relevant papers that could provide background or related work for me to read more.''} P9 appreciated \textit{``the ability to lay out the steps when I start something. [It] was very helpful [for] visualizing the outcome.''} \textbf{Second}, participants wanted the AI agent to have access to the broader context that already existed in their project documents. When using the probe, P2 was concerned that they would need to \textit{``prompt [the chatbot] with a lot of background,''} but realized they \textit{``might have [the background] written down''} in their document. 
P3 desired closer document integration but warned that the agent may clutter their document. They suggested: \textit{``having something on the side that does not flow into where I am writing, like a side [panel]---if I want something from it, I want to bring it back into my doc.''}

\subsubsection{\textbf{[F2] Literature search and understanding play a central role in research planning}}
\label{s:formative-planning}

In all project documents, we saw places where participants intended to initiate a multi-step plan (often via a \textit{how?} and \textit{can we?} question) to address different research tasks. The most common category of tasks are ones that were literature-augmented (e.g., literature search and understanding), which are often combined with a wide variety of other research actions, such as experiment/artifact design (P3), experiment/code execution (P4), data inspection and synthesis (P7), communicating and discussing results (P5), and ideation (P2). Literature was not only a significant component of \textit{existing} plans, but was important for informing \textit{future} plans. Some participants' plans (e.g., P3, P4, P7) were formed from hypotheses they hoped to verify, which prior work may have already addressed; P7 pointed out that the hypotheses they had in mind can \textit{``actually be proven from previous papers''} so they saw literature review as a planning aid. Consequently, all participants expressed a desire to use AI to help with \textit{literature-augmented tasks}---exploring and understanding relevant literature to inform decision-making. This finding echoes recent surveys on how researchers leverage LLMs to conduct research, where the most frequent usage category was information-seeking \cite{liao2024llms}.

\subsubsection{\textbf{[F3] Participants preferred to perform higher-level reasoning and synthesis themselves}}
\label{s:formative-user-steps}
In addition to highlighting tasks where they wanted AI assistance, participants also spoke about tasks they did not want AI to automate away. In particular, participants saw higher-level reasoning and information synthesis as critical tasks they wanted to do themselves. They were also often unsatisfied with AI's outputs on these tasks. P3 explains: \textit{``this tinkering process around reading and playing around with things is what gives you the ideas. I don't know if I want those things automated because the process is as helpful as the final result.''}
P5 agreed that they \textit{``wouldn't want it to be making the final decisions for sure. Just give me inspiration for where I can go.''} Specifically, they shared that \textit{``the main thing I worry about is feeling enough ownership}'' if AI makes more consequential decisions. An overeager AI assistant that attempts to perform tasks researchers prefer to do themselves is frustrating because it does not complement the user's work and instead creates undesired noise for them to filter through.

\subsection{Design Goals}

We synthesize our formative study findings into three design goals for an interactive system that facilitates meaningful collaboration between researchers and AI agents by using plans as a shared representation.

\begin{itemize}
    \item[\textbf{DG1:}] \textbf{Provide opportunities for human-agent collaboration across planning and execution.} 
    Agent developers have called for decoupling planning and execution to avoid fruitless execution \cite{chip-agents}. Indeed, participants discussed \textit{planning} (i.e., breaking down a problem into smaller, more actionable tasks like they did in their project documents) and execution (i.e., completing steps in the plan and refining outputs, with or without AI assistance) as separate but complementary actions. 
    We thus see planning and execution as two distinct stages for fertile researcher-AI collaboration. Because these two stages are closely intertwined and synergistic, we aim to interleave the two in our system.

    \item[\textbf{DG2:}] \textbf{Allow flexible delegation of work between researcher and agent.} Researchers may not want to delegate all parts of a plan to AI [F2, F3]. 
    We also observed that this preference may also be context-dependent and constantly shifting. Our system uses a concrete implementation of a theory of flexible delegation proposed by Satyanarayan \cite{satyanarayan2024intelligence}.

    \item[\textbf{DG3:}] \textbf{Integrate seamlessly into a document environment.}
    Project documents contain rich externalizations of researchers' workflow, thought processes, and plans (e.g., ``to-dos'') [F1]. This is key information an AI agent can use to better assist the researcher. The document is also an appealing environment for researcher-agent collaboration, but requires careful information management strategies to prevent unwanted distractions [F3].
    Thus, we aim to 1) allow the researcher to use familiar document editing affordances to interact with the agent, and 2) strategically manage outputs to avoid excessively cluttering the document.
     
\end{itemize}

\section{Cocoa: System Walkthrough and Implementation}
\label{s:system}

We present \system, a system that embeds an AI agent into a document editor and provide an affordance we call 
\textit{interactive plans} that allow users to collaboratively plan (\textbf{co-plan}) with the agent---the agent proposes an initial plan of execution to tackle a user request that the user can edit to their liking. Then, users can collaboratively execute (\textbf{co-execute}) the plan with the agent---the user and the agent can build off each other's work to synergize human and AI capabilities, and provide flexible delegation of human and AI work. 

To illustrate the features and functionality of \system, we follow the journey of \name, a fictional researcher in human-AI interaction, as she uses her project document to further explore open questions in interactive interfaces for AI agents.

\subsection{Co-Planning}
\label{s:co-planning}

\name's project document is an informal, reverse chronological log of research progress and includes information such as meeting notes, rough ideas, links to literature, and questions for herself and her collaborators. She has a specific question in her document that she wants to explore further, but being new to the topic, she is not sure how to proceed on her own: \textit{``What are new ways AI agents can interactively elicit human feedback?''} 

\subsubsection{Reviewing and selecting an initial plan}

\begin{figure*}
    \centering
    \includegraphics[width=0.9\linewidth]{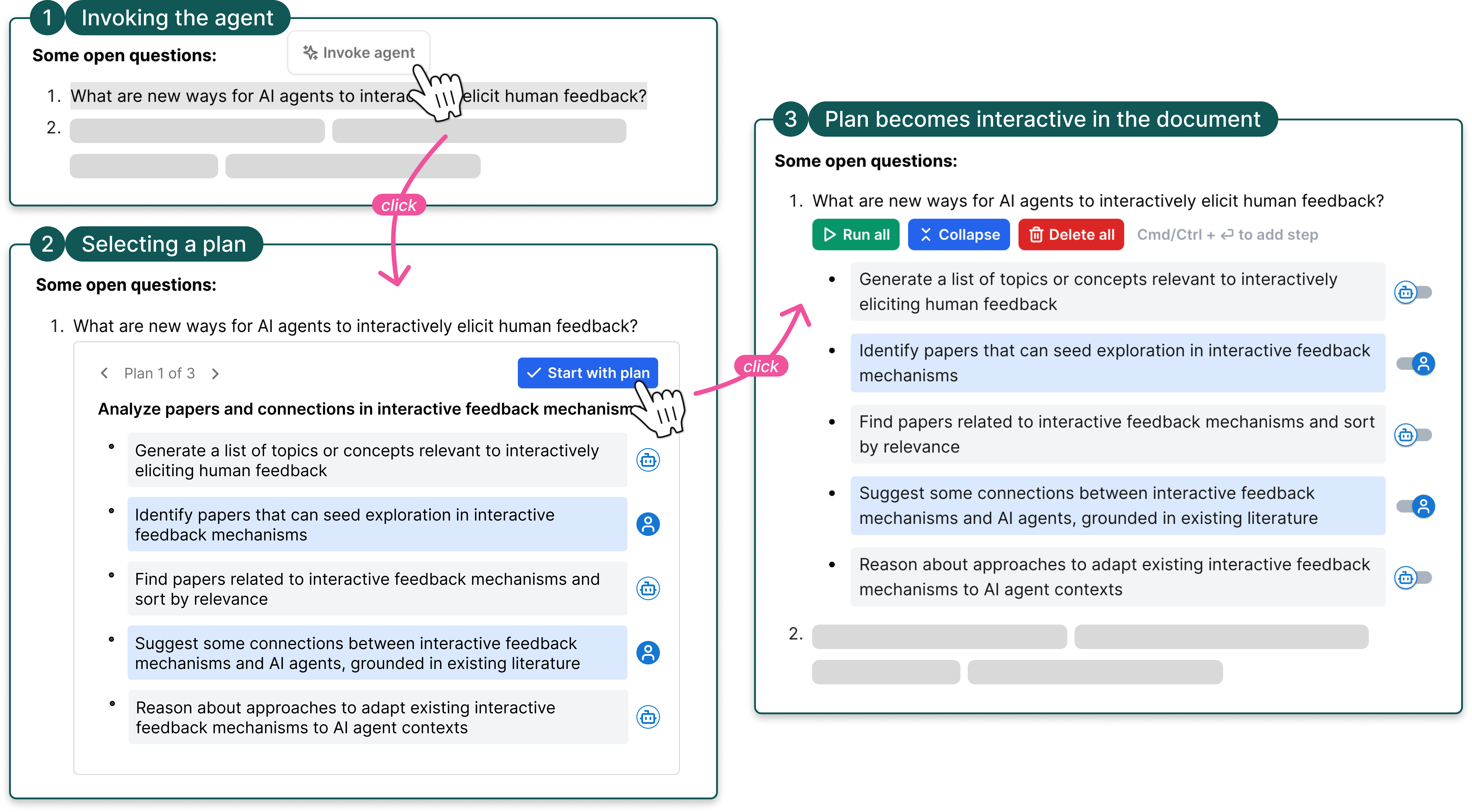}
    \caption{A user invokes the agent on a piece of text in the document by clicking on the ``Invoke agent’’ button that appears on hover whenever text is highlighted. The agent will use the highlighted text and context from elsewhere in the document to propose a series of plans, displayed in a plan selector that appears under the highlighted text. Once the user selects a plan, it becomes fully interactive in the document.}
    \label{fig:agent-start}
    \Description{A user invokes the agent on a piece of text in the document by clicking on the ``Invoke agent’’ button that appears on hover whenever text is highlighted. The agent will use the highlighted text and context from elsewhere in the document to propose a series of plans, displayed in a plan selector that appears under the highlighted text. Once the user selects a plan, it becomes fully interactive in the document.}
\end{figure*}

To initiate co-planning, she highlights the question to invoke the agent on it using a button that appears in a floating menu. The agent analyzes the question within the context of her entire document and proposes a few different plans via a \textbf{plan selector} UI. Noticing one of the proposed plans started with generating a list of more detailed concepts, which would potentially give her the opportunity to learn and tune the specific paper search directions, \name selects the plan, which then becomes fully interactive within the document (Figure~\ref{fig:agent-start}). 

\subsubsection{Assigning ``Agent steps'' and ``User steps''}
The initial plan included steps for brainstorming relevant topics, searching for papers, and making connections between papers (Figure~\ref{fig:agent-start}). 
To balance between effort and control, \name uses the \textbf{step assignment toggle} (Figure \ref{fig:teaser-detailed} B\&C, Figure \ref{fig:agent-start} right) to assign tasks that are relatively low risk but high effort to the agent, such as searching for papers and expanding on some of her preliminary ideas. In contrast, she assigned tasks that were more consequential or required higher-level thinking to herself, such as identifying seed papers for the agent to explore more broadly or making insightful connections between multiple papers.

\subsubsection{Plan Editing and Interactive Replanning}
\begin{figure*}
    \centering
    \includegraphics[width=0.9\linewidth]{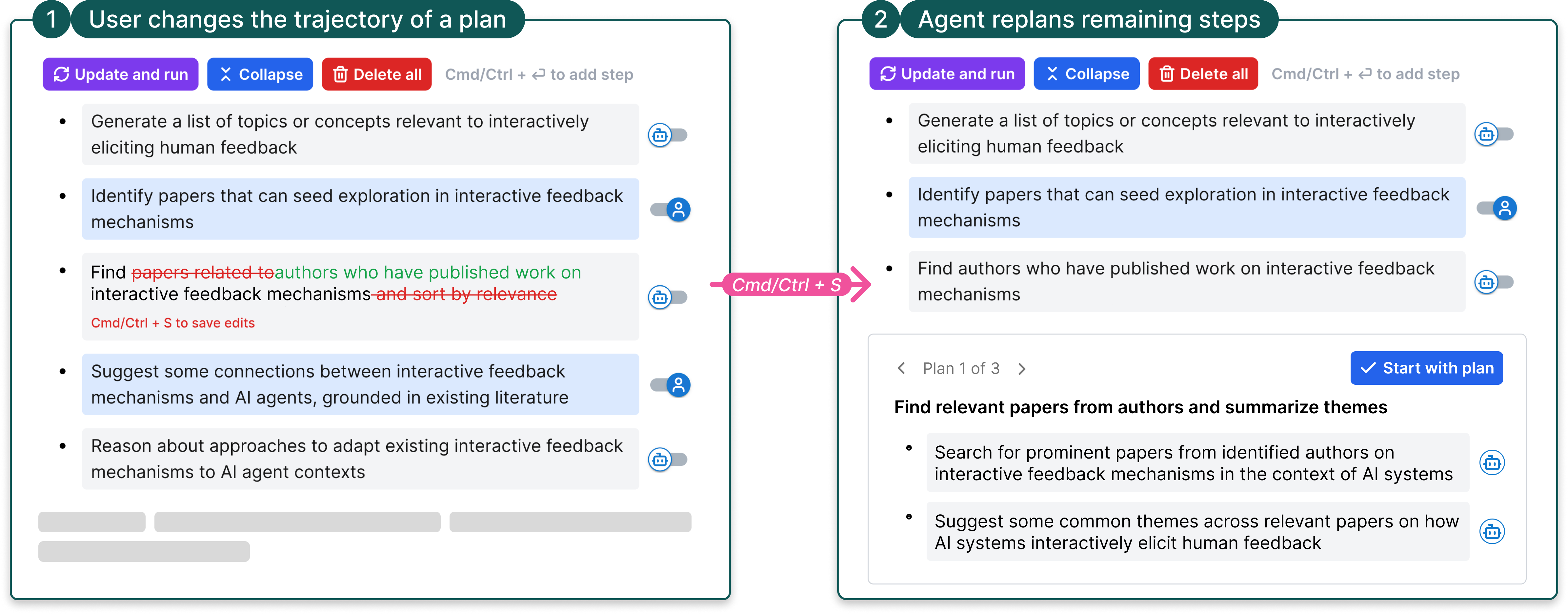}
    \caption{If major changes are made to a step that changes the rest of the plan’s trajectory, \system detects this and will trigger replanning. Replanning replaces subsequent steps with ones the agent suggested for ``autocompleting’’ the plan.}
    \label{fig:replanning}
    \Description{The user has edited a step significantly, such that it changes the rest of the plan’s trajectory. Cocoa detects this and replaces subsequent steps with ones the agent suggested for autocompleting the plan.}
\end{figure*}

Upon further review of the plan, \name is now also curious about the prominent authors in the area and wish to learn about their papers first. She edits the \textbf{step description} to reflect this. The step description can be edited just like editing any other bulleted list in her document, making this intuitive for \name. 
Similarly, \name can also easily \textbf{add and delete steps from the plan}. For example, she notices that the agent has proposed a summary step that may be irrelevant, so she highlights the step and hits \texttt{Backspace} to delete it. Finally, she uses the keyboard shortcut \texttt{Cmd/Ctrl+Enter} while selecting the last step to add a new step below it, and edits the step description of the newly added step 
These interactions mirror how one would edit items in a bulleted list in text editors, making them intuitive for \name.  
While \system allowed \name full control over how she wishes to manually edit the plan, this process can be effortful for users. Therefore, the system provides support for \textit{replanning} in the following ways:  
\textit{First}, when editing an individual step, \name can request the system to provide suggestions for alternatives, informed by the previous steps and document context. 
\textit{Second}, after the user has edited some plan step, the plan may become incoherent. For example, a plan step that searches for additional relevant papers given a set of seed papers might no longer make sense if the previous step were edited to produce a list of prominent authors. In this case, \system \textbf{automatically detects this and replans subsequent steps} accordingly by removing those steps and ``autocompleting'' the plan with new ones (Figure \ref{fig:replanning}).

\subsection{Co-Execution}
\label{s:co-execution}

Now that \name has made her desired edits to the plan, she is ready to co-execute it with the agent.

\subsubsection{Continuous and stepwise execution}
Drawing inspiration from computational notebooks, there are two modes of co-execution: continuous and stepwise. Each step has a \textbf{step completion indicator} that indicates whether the step has not yet been run (a bullet point), is in progress (a spinner), requires user input (a question mark badge), or is complete (a checkmark). This notebook-inspired design 
allows \name to dispatch the agent to run autonomously on parts of a plan she trusts the agent to complete on its own. She can also iterate on specific parts of a plan by examining and potentially amending the output of each plan step before continuing. 

\name wishes to run through the entire plan once to see the output. She triggers \textit{continuous co-execution} with the \textbf{[Run all]} button at the top of the plan. In \textbf{continuous co-execution}, the agent will automatically continue onto the next step as soon as it (or the user) has completed the previous one, until it reaches a user step or the end of the plan. 
By contrast, if she wishes to iterate and refine the outputs of individual steps before the agent moves onto the next one, she can opt for \textbf{stepwise co-execution}.

\name can flexibly switch from stepwise to continuous co-execution at any point by clicking the \textbf{[Run remaining]} button at the top of the plan. Conversely, \name can switch from continuous to stepwise co-execution by clicking on the \textbf{[Pause after this step]} button during execution and manually running subsequent steps.\footnote{\system does not permit out-of-order stepwise execution because some steps may rely on the outputs of previous ones. We also note that out-of-order execution in computational notebooks is a major user pain point identified in prior academic work \cite{Chattopadhyay2020WhatsWW, lau2020notebooks} and industry reports \cite{jupyter2015survey, grus2018notebooks}.} 

\subsubsection{Editing outputs of agent steps}
\begin{figure*}
    \centering
    \includegraphics[width=1\linewidth]{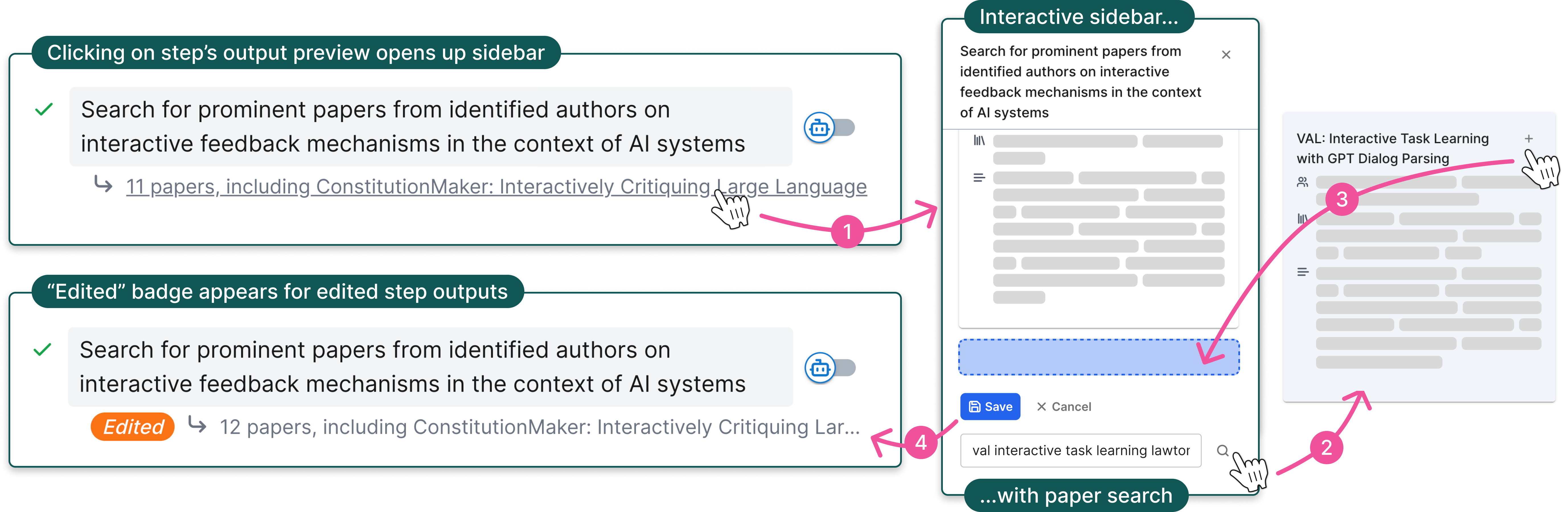}
    \caption{Users can click the output preview text beneath the plan step to access the interactive sidebar (1). By editing agent outputs in the interactive sidebar (2), users draw from their expertise to edit outputs, such as adding papers the agent might have missed (3), to guide the agent. Once edited, an indicator will appear in the output’s text preview below the step (4).}
    \label{fig:output-editing}
    \Description{A user clicks the output preview text beneath the plan step to access the interactive sidebar that contains some card UIs displaying paper information. Users can delete card UIs they consider irrelevant. By editing agent outputs in the interactive sidebar, users draw from their expertise to guide the agent. Once edited, an indicator will appear in the output’s text preview below the step.}
\end{figure*}

As the agent completed a paper search step, \name clicks into that step to examine its output. This opens a sidebar to the right that is populated with interactive paper cards (Figure \ref{fig:output-editing}). \name can further guide the agent with her expertise by removing papers she considers irrelevant and/or adding papers the agent missed using a built-in paper search functionality connected to the Semantic Scholar academic database \citep{Kinney2023TheSS}. For now, she removes a couple of papers in the output that were not published recently and were published in a less relevant venue.

The interactive UI in the sidebar dynamically adapts according to the step's \textit{output type}, which falls into one of five categories in \system: \textbf{papers}, \textbf{authors}, \textbf{topics}, \textbf{entities}, and \textbf{text}. Papers, authors, and topics are drawn from Semantic Scholar and are displayed as interactive cards that \name can add, delete, or search for using the sidebar's built-in Semantic Scholar search. Entities are lists of items in natural language (search queries, research questions, etc.) and are displayed as editable pills. Text is natural language text displayed in an editable text box.

If the agent fails to complete its step (e.g., it finds no papers or runs into an error), \system will also proactively alert \name so that she can complete the step instead, or retry with the agent. 

\subsubsection{Completing user steps}
Before executing the plan, \name had assigned step 2 to herself during co-planning. She wanted the agent to start a broad search of many relevant topics and concepts, but then wanted to contribute some seed papers she knew would be relevant to steer the agent's subsequent actions. 

\system has now paused at step 2, highlighting the step in orange to attract \name's attention and to seek her input. When she clicks into that step, the same interactive sidebar opens to the right (Figure \ref{fig:output-editing}). She could then manually review the topics generated in step 1 and uses the paper search feature built into the sidebar to browse relevant papers and add them to the output of step 2.
Additionally, she recalls a few papers that a collaborator had recommended to her previously, finds them using the paper search feature, and adds them to the output.

\subsubsection{Final output panel}
\begin{figure}
    \centering
    \includegraphics[width=1\linewidth]{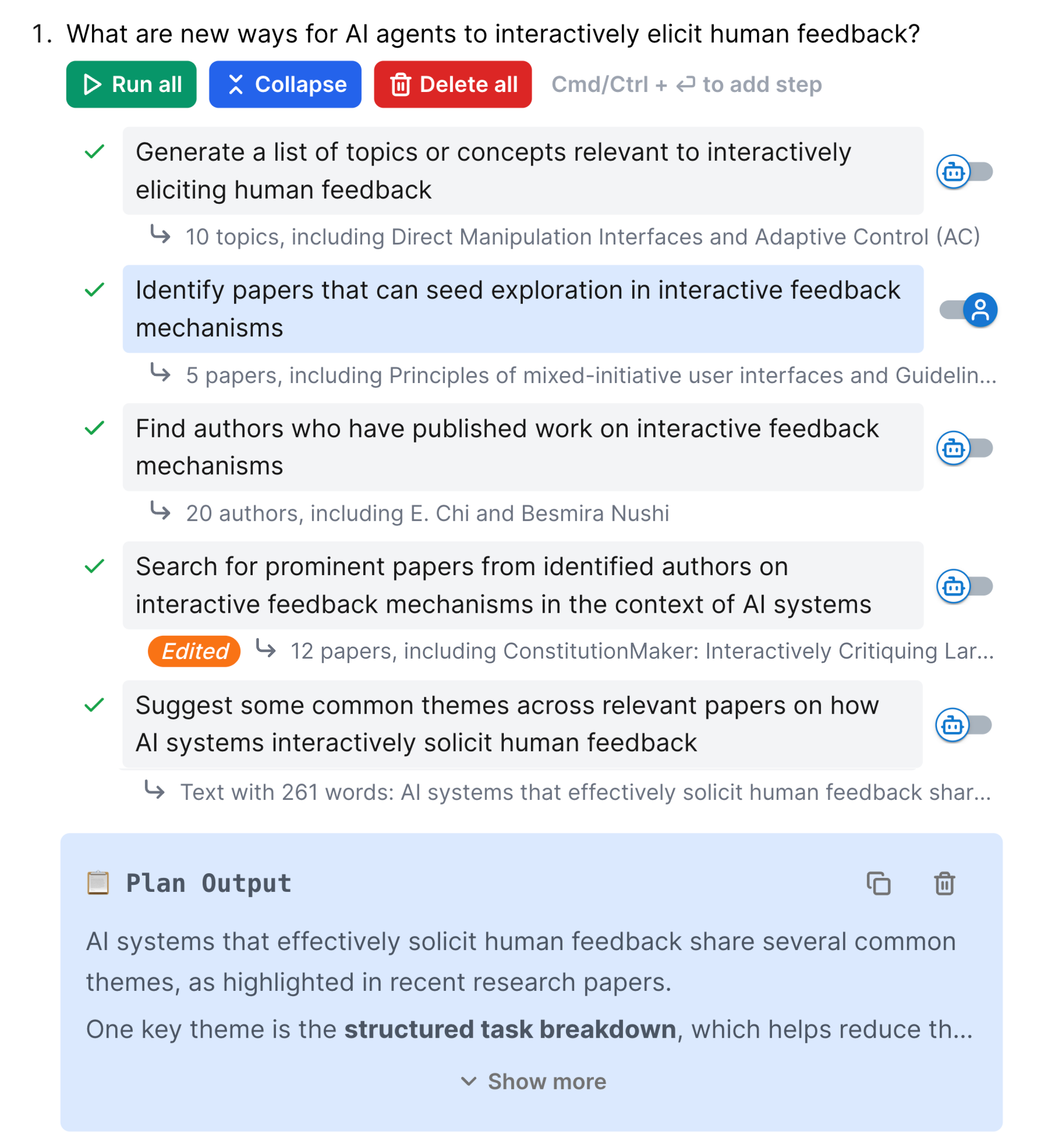}
    \caption{Once execution of the entire plan is complete, the agent inserts a plan output panel in the document with the last step’s output. This allows the user to view the plan’s results within the context of their document. The panel can be expanded or collapsed, and remains in the document even if the plan itself is collapsed for easy access.}
    \label{fig:plan-output}
    \Description{A light blue panel appears at the bottom of a plan once it's finished executing to allow the user to view the plan’s results within the context of their document. Content from the panel may be copied by the user and pasted elsewhere in the document. The panel can be expanded or collapsed, and remains in the document even if the plan itself is collapsed for easy access.}
\end{figure}

After the final plan step has been executed, the agent adds a \textbf{plan output panel} (Figure \ref{fig:plan-output}) containing a modified version of the last step's output into the document just below the plan. The modification involves rewriting the output, given the context of the plan and its outputs, to more directly connect to the original user request. This brings the results of running the plan into the document so \name can easily reference it in the context of other content. The panel is collapsed by default to save space and reduce clutter, but she can expand it to copy and paste parts of the output she finds useful and add them to her document. If she does not want the panel in her document at all, she can simply delete it; she can still access the outputs of all steps in the sidebar.

\subsection{Interleaving Co-Planning and Co-Execution}
In addition to supporting co-planning and co-execution independently, \system is designed to support \textbf{synergistic interleaving of the two}. As \name co-executes the plan with the agent, she encounters a relevant author that the agent identified and wishes to further explore the author's papers. She edits the following step's plan description to satisfy this. She reruns that step and receives a new batch of papers, and the agent automatically reruns the rest of the plan with the updated output. She is also not satisfied with the agent's interpretation of common themes and believes that she can surface deeper insights by reading the papers herself, so she toggles that step to be a user step. After reading the papers in detail, she returns to her document and jots down her insights. 
She now finds the next step repetitive, so she deletes it. 
Overall, her use of \system is highly iterative, and she smoothly transitions between co-planning and co-execution. 

When designing \system, we were inspired by the computational notebook paradigm supported scientists via tight coupling of code and output, enabling rapid iteration and steering of complex scripts. To summarize this connection, we map \system's features to those in computational notebooks in Table \ref{t:cocoa-notebook}.

\begin{table*}[h]
\small
\centering
    \begin{tabular}{p{8.5cm} p{8.5cm}}
    \toprule 
    \system Feature & Computational Notebook Feature \\
    \midrule  
    Editable plan steps that can be composed into a larger, more complex task. & Code cells that can be composed into a larger, more complex program. \\
    Users can choose between stepwise and continuous execution. & Users can choose between running individual cells or run all. \\
    Users can view and improve step outputs within the plan. & Users can view code outputs within the notebook. \\
    Users can flexibly switch between improving the plan and executing the plan steps. & Users can flexibly switch between improving the code and executing code cells. \\
    Plan steps assigned as user steps pause execution so users can perform tasks themselves. & Break points pause execution so users can examine current code outputs; all cells are executed programmatically. \\
    Completed plan can serve as documentation of agent \& human work. & Completed notebook can serve as documentation of analyst's work. \\
    \bottomrule
    \end{tabular}
    \caption{Main connections between features in \system and those in computational notebooks.}
    \label{t:cocoa-notebook}
    \Description{A 2-column table with headings Cocoa Feature and Computational Notebook Feature. There are 3 rows, where each row contains a description of the feature in Cocoa and computational notebooks.}
\end{table*}

\subsection{Implementation Details}
\label{s:cocoa-implementation}

\system is implemented as a web application with a Next.js and TypeScript frontend communicating with a Flask backend. The frontend uses the Tiptap\footnote{https://tiptap.dev/} framework for the main document editor. Each document supports synchronous collaboration via Hocuspocus\footnote{https://tiptap.dev/docs/hocuspocus/} and all changes are auto-saved to Tiptap Cloud, which also serves as storage for all participant documents. Interactive components within the document editor are implemented as custom Tiptap extensions in TypeScript. 

The Flask backend orchestrates LLM activity with calls to a custom tool-calling LLM agent\footnote{\url{https://platform.openai.com/docs/guides/function-calling}} powered by GPT-4o and scaffolded by LangChain.\footnote{https://www.langchain.com/} Our agent is designed to assist researchers through literature-grounded response generation. To do this, we implemented tools that give our agent access to millions of research papers via publically available Semantic Scholar APIs.\footnote{https://www.semanticscholar.org/product/api} This API is powered by The Semantic Scholar Open Data Platform \cite{Kinney2023TheSS} which consists of a state-of-the-art PDF extraction and knowledge graph normalization pipeline. This gave our agent access to both the full text of millions of research papers and also their rich metadata, such as authors, fields of studies, and publication venues \cite{Kinney2023TheSS}. 
More implementation details can be found in Appendix \ref{a:implementation}.

\section{Lab Study}
\label{s:lab-study}

We conducted a within-subjects task-based lab study with 16 researchers to evaluate \system against a custom chat baseline. Specifically, our study aimed to address the following research questions:

\begin{itemize}
    \item[\textbf{L1:}] How does \system compare to our chat baseline for ease of use, steerability, and general utility\footnote{General utility is a composite metric created from Q1, Q2, and Q4 from Appendix \ref{a:eval-likert-form}.} in research project documents?
    \item[\textbf{L2:}] How does interleaved co-planning and co-execution contribute to ease of use, steerability, and general utility in \system, if there is a significant improvement over the baseline?
    \item[\textbf{L3:}] What kinds of steps did researchers wish to assign to an agent and themselves in practice?
\end{itemize}

\subsection{Participants}
We recruited 16 Ph.D. and postdoctoral researchers (14 Ph.D.s, 2 postdocs; 10 female, 6 male) in computer science (CS) or CS-adjacent areas via university mailing lists, word of mouth, social media recruitment messages (on Twitter/X, Mastodon, Bluesky), and personal connections. We recruited on a first-come, first-serve basis as we conducted our studies and closed recruitment when we approached data saturation. Further details about our recruitment process, pilot studies, and participants can be found in Appendix \ref{a:user-study-recruitment} and Table \ref{t:participants-eval} of Appendix \ref{a:user-study-participants}.

\subsection{Baseline System}
\begin{figure*}
    \centering
    \includegraphics[width=0.9\linewidth]{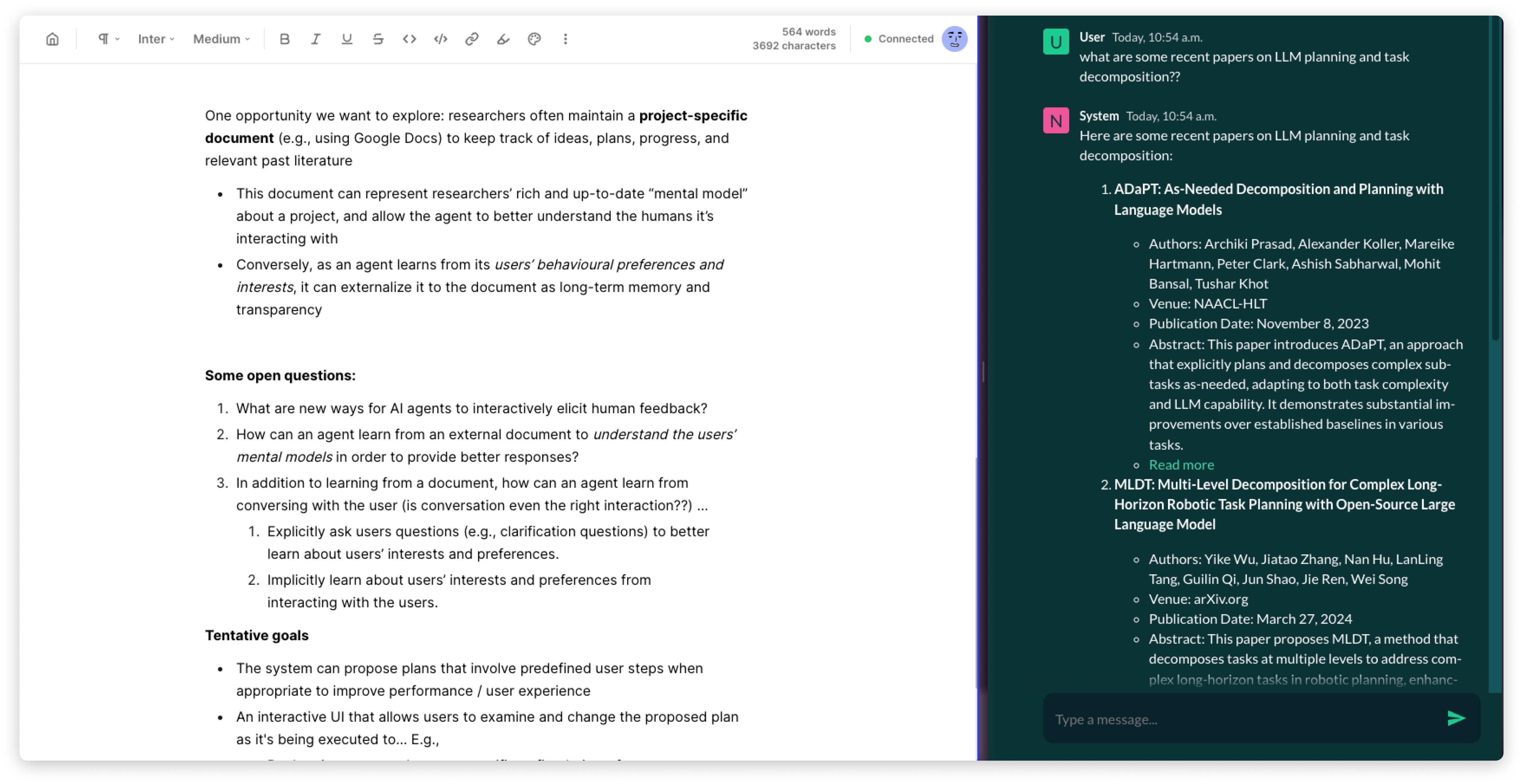}
    \caption{Participants' typical setup for our baseline condition. On the left, the same document editor from \system but without an option to invoke the agent for co-planning and co-execution. On the right, the agent is situated within a chat interface.}
    \label{fig:baseline-setup}
    \Description{Participants' typical setup for our baseline condition. On the left, the same document editor from Cocoa but without an option to invoke the agent for co-planning and co-execution. On the right, the agent is situated within a chat interface.}
\end{figure*}

As our primary research question is to better understand the costs and benefits of \system{}'s co-planning and co-execution affordances, the baseline system (Figure~\ref{fig:baseline-setup}) was a common chat interface powered by the same underlying LLM agent: a custom tool-calling agent powered by GPT-4o with access to the same state-of-the-art tools for searching, analyzing, summarizing, and comparing literature across millions of papers on Semantic Scholar \cite{Kinney2023TheSS}, in addition to standard LLM capabilities. See \ref{a:implementation} for more details about the agent. 
The design of the chat interface closely resembled that of existing AI-powered research systems (e.g., Assist mode in Scite \cite{scite}, Google Scholar Labs, various deep research systems), and popular LLM chatbots (e.g., ChatGPT, Claude). During the study, participants positioned the chat interface beside the baseline document editor, which is the same editor that \system uses, except we removed the option to invoke the agent to eliminate all co-planning and co-execution interactions. 

We chose this baseline because it mimicked the dominant interaction paradigm implemented in AI-powered research tools at the time of our study and allowed for strong comparisons on aspects like ease of use. We recognize that variations of this baseline, such as forcing the agent to show an editable plan before execution (e.g., \cite{he2025plan}), are also strong baselines that can help further isolate the factors contributing to our results. We discuss ablations \system as alternative baselines in Section \ref{s:limitations-fw}. 

\subsection{Research Task and Study Procedure}

To make the study more realistic and grounded in real-world research projects, tasks for this study were open questions or unfinished items from participants' existing project documents. Details for how we controlled for the length of the study and ensured the two tasks were valid and comparable can be found in Appendix~\ref{a:user-study-details}. Tasks were randomly assigned to \system and the baseline. 

The first author (study facilitator) conducted 1:1 studies with the 16 participants over Google Meet between October and December 2024. Each study was 90 minutes in length. The study started off with introductions and a brief overview, followed by the two tasks, which were counterbalanced across participants. Detailed procedures for each condition can be found in Appendix \ref{a:user-study-details}. After each condition, participants filled out a short evaluation form with 5-point Likert scale questions for that condition (see Appendix \ref{a:eval-likert-form}). These questions were inspired by the System Usability Scale \cite{brooke1996sus} but adapted for the research context.

The task-based portion of the study was 75 minutes. The study concluded with a semi-structured interview that lasted around 15 minutes. Participants were asked to reflect on their experiences across the two systems and discuss the pros and cons of both. They were also asked about particular decisions the study facilitator observed when using each system. 

Participants received a \$75 USD honorarium upon completion of the study. All studies were recorded and transcribed by Google Meet. This study was reviewed and approved by the Allen Institute of AI's IRB.

\subsection{Data Analysis}
\label{s:data-analysis}
The first author analyzed the recording transcripts using Braun and Clarke's reflexive thematic analysis \cite{Braun2019ReflectingOR}. This approach uses a hybrid inductive-deductive approach to iteratively surface codes and themes across the data. We paid close attention to codes related to participants' comparisons of their experiences across the two systems and gradually grouped them into themes. NotebookLM\footnote{https://notebooklm.google.com} was used to help iterate on themes and discover new ones. We performed statistical tests of participants' ratings on the evaluation forms with the Wilcoxon signed-rank test (with the Bonferroni correction for our multi-rating analysis on significant results) given the non-parametric nature of our data. We saved all documents used in the study as well as all conversation histories in the baseline system, and referred back to them when needed.

The first author also coded all video recordings from the study to note the timestamps at which participants engaged and disengaged in a particular interaction. In \system, these interactions were \textbf{co-planning} (see \ref{s:co-planning}), \textbf{co-execution} (see \ref{s:co-execution}), and \textbf{output inspection} (when the participant passively inspected the system's outputs). The closest equivalent interactions were used as codes in the baseline: \textbf{prompting for planning} (instructing the agent to perform an action without any attempts to modify the agent's earlier outputs), \textbf{prompting for editing}\footnote{Since there were no user steps in the baseline, the closest equivalent interaction is attempting to edit the agent's output via prompting.} (instructing the agent to modify or only consider a subset of its output in future actions), and \textbf{output inspection}. 

We established these equivalences based on the functional purpose of the interactions: co-planning and prompting for planning both instruct the agent to produce new outputs, while co-execution and prompting for editing aim to improve an existing output. We note that there is not perfect correspondence between interactions across the two systems. For example, co-execution is not only about refining agent outputs, but also giving users agency to complete tasks they assigned for themselves. We acknowledge this by shading the interactions in \system and the baseline with slightly different colours in Fig. \ref{fig:timestamps-vis}. However, because co-planning and co-execution subsume the functional roles of their baseline counterparts in the context of our study, we consider them appropriate to compare.  

\section{Lab Study Results}
\begin{figure*}
    \centering
    \includegraphics[width=1\linewidth]{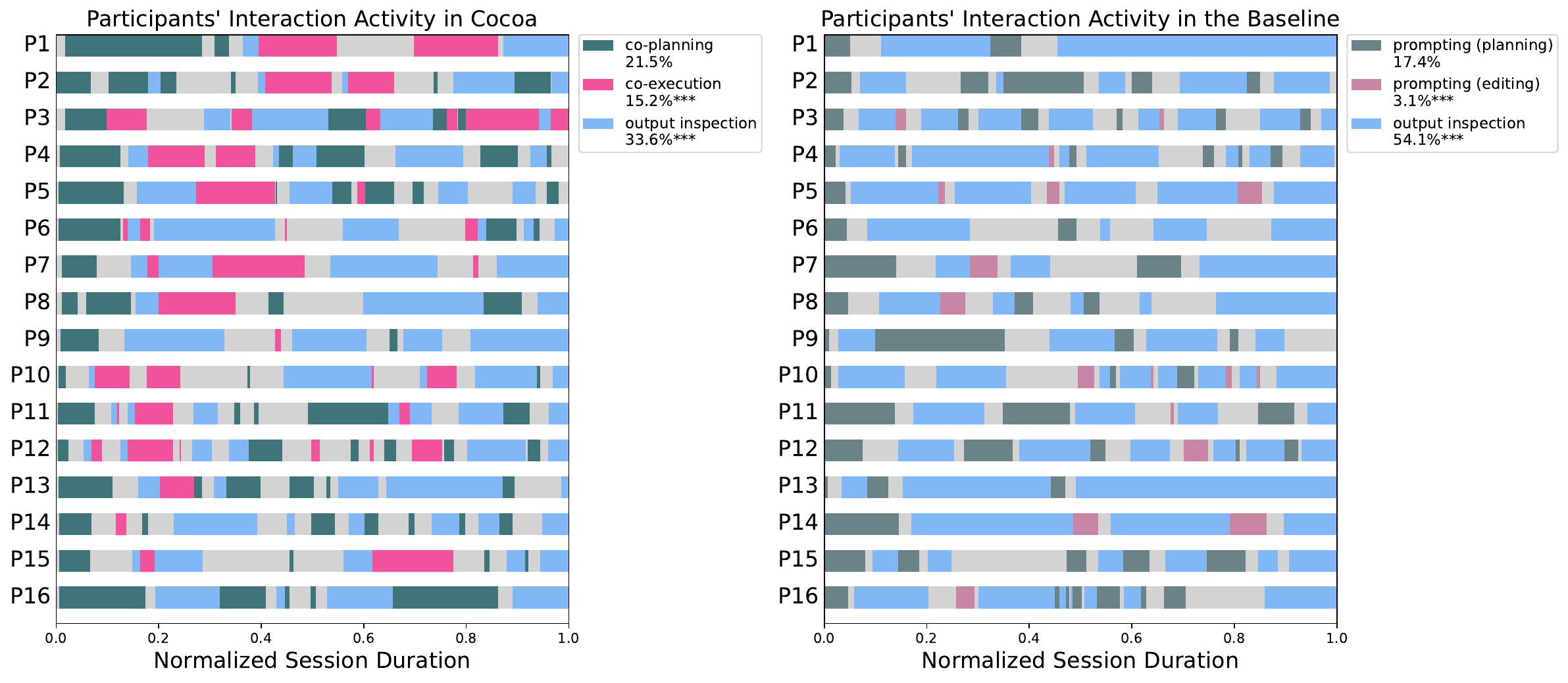}
    \caption{Participants' interaction activity in \system and the baseline within the study sessions. *** indicates a statistically significant difference ($p < 0.001$ via a paired t-test) with the closest equivalent interaction in the opposing condition. Gray areas represent spans of time where the participant was not engaging in any of our coded interactions because they were answering a question posed by the study facilitator and/or waiting on the system. Definitions for all coded interactions can be found in Section \ref{s:data-analysis}.}
    \label{fig:timestamps-vis}
    \Description{A visualization of participants' interaction activity in Cocoa (left) and the baseline (right). Each participant on the y-axis is represented as a bar that is divided into chunks of co-planning, co-execution, and output inspection for Cocoa, and prompting (planning), prompting (editing) and output inspection for the baseline. The x-axis depicts normalized session duration from 0.0 to 1.0.}
\end{figure*}

A high-level overview of participants' interactions with \system and the baseline is depicted in Figure \ref{fig:timestamps-vis}. In both conditions, output inspection occupied the most time out of the coded interactions, but was lengthier in the baseline than \system (54.1\% vs. 33.6\% of the approximately 25-minute session). A paired t-test found this difference to be significant ($t(15) = 3.95, p < 0.001$). Participants also engaged in co-execution in \system significantly more often than the closest equivalent interaction---output editing via prompting---in the baseline (15.2\% vs. 3.1\%, $t(15) = 4.29, p < 0.001$). They spent slightly more time on co-planning (21.5\% vs. 17.4\%), although this was not significant ($t(15) = 1.01, p = 0.33$). Our analysis indicates that, when provided with more affordances to interactively modify the agent's outputs, users will take advantage of those opportunities for active engagement and shift away from passive output consumption. That is, users could \textbf{interleave co-planning and co-execution to steer the agent through tighter feedback loops}, rather than relying on slower cycles of trial-and-error prompting and output inspection.

We also computed performance metrics for both co-planning and co-execution from coded interactions and system logs. These included the number of step edits, reassignments, and output edits across all participants. Table \ref{t:usage-metrics} displays our results. Overall, 32.6\% of steps were edited and 17.4\% were reassigned. This shows that most plan steps proposed by the agent (the content of the step as well as the assignment) were accepted as-is. 
During execution, the agent surfaced an average of 12.9 discrete output items (i.e., papers, authors, topics, text entities) per step to users. On average, 2.1 (16.3\%) were deleted by participants, suggesting that the system had reasonable precision. 

However, these metrics alone may not paint a holistic picture about \system's use. To more deeply understand the context and rationale behind users' actions, we analyzed videos from our user studies (Fig. \ref{fig:timestamps-vis}) alongside qualitative transcripts from participants thinking aloud (Section \ref{s:lab-replanning}). 

\begin{table}[h]
\centering
    \begin{tabular}{p{5cm} p{2.5cm}}
    \toprule 
    Metric & Count (\%, if\newline applicable) \\
    \midrule  
    Total steps completed across all plans & 86 \\
    Step edits by user & 28 (36.5\%) \\
    Step reassignments by user & 15 (17.4\%) \\
    Step additions by user & 12 (14.0\%) \\
    Step deletions by user& 4 (4.7\%) \\
    Replanning triggered by system& 14 \\
    Avg. output items (i.e., papers, authors, topics, text entities) per step & 12.9 \\
    Avg. output items deleted per step & 2.1 (16.3\%)\\
    \bottomrule
    \end{tabular}
    \caption{\system usage metrics from our lab study.}
    \label{t:usage-metrics}
    \Description{A 2-column table with headings Metric and Count. There are 8 rows, where each row contains a metric with its corresponding count and percentage, if applicable.}
\end{table}

In the rest of this section, we present quantitative and qualitative analyses of participants' data from our user study. We group these results by our research questions from the start of Section \ref{s:lab-study}. We denote the median rating of our baseline and \system as $M_b$ and $M_c$, respectively, and the Wilcoxon test statistic as $W$.

\begin{figure*}
    \centering
    \includegraphics[width=1\linewidth]{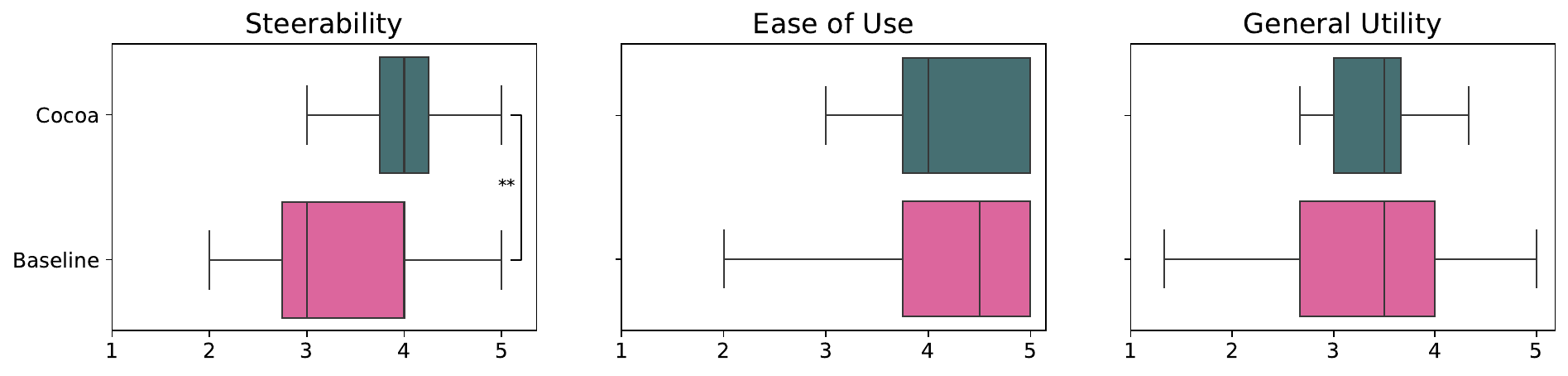}
    \caption{Participants' Likert scale ratings of steerability, ease of use, and general utility across our baseline and \system. ** indicates $p < 0.01$ via a Wilcoxon signed-rank test.}
    \label{fig:cocoa-boxplots}
    \Description{Three box plots showing participants' Likert scale ratings of steerability, ease of use, and general utility across Cocoa and the baseline. Cocoa is shown above the baseline on the y-axis in all 3 plots. All 3 plots have an x-axis depicting the Likert scale rating from 1 to 5. Participants rated Cocoa's steerability as significantly higher compared to the baseline.}
\end{figure*}

\subsection{Steerability, Ease of Use, and Utility (L1)}
\label{s:likert-results}

Introducing novel interactive systems with additional affordances can often lead to higher effort when using the systems as a trade-off for better utility \cite{Shneiderman1997DirectMV, Hutchins1985DirectMI}. However, based on participants' post-task ratings on a 5-point Likert scale, there was no significant difference in perceived \textit{ease of use} between the baseline and \system with 16 participants ($M_b = 4.5, M_c = 4, p=1.000, W=7.50$). 
At the same time, we observed a significant difference that \system provided better \textit{steerability}. In response to the question \textit{``I could easily steer the system towards doing something helpful,''} participants rated \system higher than the baseline to a significant degree ($M_b = 3, M_c = 4, p=0.005, W=0$),\footnote{Here, we set the significance threshold to $\alpha = 0.05 / 3 = 0.015$ using Bonferroni correction. Our result is significant post-correction: $p=0.005 < 0.015$.} with $W=0$ indicating that all participants rated \system's steerability as greater than or equal to that of our baseline. 
In sum, these results suggest that \textbf{\system affords interleaving co-planning and co-execution via rich interactive affordances without sacrificing ease of use}, when compared to the simple chat interface used in the baseline condition (Figure~\ref{fig:cocoa-boxplots}).

We also tested for differences in \textit{general utility}, which was a composite metric consisting of three measures broadly related to the usefulness and insightfulness of system outputs, but we found no significant differences ($M_b = M_c = 3.5, p=0.7, W=40$). This was unsurprising---the systems' utility to the researcher may depend on a range of system-agnostic factors beyond our control (e.g., contextual interpretations of outputs by researchers, the extent to which researchers thought about the task before the study, etc.) However, improved steerability may have allowed users to steer the agent away from providing downright unhelpful outputs in \system, as indicated by fewer lowly-rated outliers (see Figure \ref{fig:cocoa-boxplots}).

\subsection{Interleaving Co-Planning and Co-Execution Created an Effective Steering Loop (L2)}
\label{s:lab-replanning}
\subsubsection{\textbf{Outputs from co-execution drove re-planning}}
Qualitative results based on think alouds and post-interviews suggest that \system was valuable not only because it enabled human-agent co-planning or co-execution, but because \textit{interleaving the two meant that co-execution outputs could be used to inform collaborative re-planning}. Rather than treating the plan as a static artifact, participants used co-execution to find interesting and potentially productive directions to steer the agent towards. For example, P7 and P15 both used outputs from agent steps as a signal to \textit{``organize thoughts and iterate on ideas''} (P7). P15 specifically iterated on an intermediate ``paper search'' step by re-executing it to cover papers from venues that they preferred. This led to subsequent plan updates that ultimately resulted in a more relevant final output. Similarly, P6 observed that ``backtracking'' to a specific step and editing it was more helpful than digging through a chat conversation: \textit{``With chat, when it does something wrong there's not really an easy way to fix it because there's no concrete steps that it's following, whereas [with \textsc{Cocoa}], I can go back and be like, let me have it do something else here.''} 

Besides modifying intermediate steps, participants also added steps at the end of the plan to assign follow-up investigations, or deleted steps to sharpen the plan's focus. For example, P12 added a step at the end of their plan to have the agent suggest some potential next steps to pursue after seeing its paper summaries, while P5 removed a less relevant intermediate step on searching for papers related to autonomous vehicles to avoid distractions from this tangential topic. In general, iterative output evaluation and re-planning allowed participants to better ``fine-tune'' the agent's outputs and also their own research process---P1 likened their experience to using \textit{``[plan steps] as building blocks''} that they could mix and match to create a custom workflow.

\subsubsection{\textbf{Fluid movement between co-planning and co-execution enabled granular steering}}
\label{s:steerability-co-execution}
The interleaving of co-planning and co-execution allowed participants to move fluidly between the two and exert control exactly when necessary. By opting for stepwise execution of the plan, participants could pause the agent to ``catch up'' with agent activity and ensure outputs are up to standard before continuing. Indeed, P4 and P11 both \emph{verified} and \emph{manually curated} the list of papers returned by the agent from a paper search step before proceeding. P11 specifically cites resource efficiency as a practical reason: \textit{``If I update [outputs of] the first or second step [after executing all steps], it would need to rerun the following steps. I just want to reduce some API usage.''} P10 similarly mentioned that they would ``waste energy'' running subsequent steps if they later detected errors in initial steps. Overall, many participants chose stepwise execution because it \textit{allowed more room to interleave co-planning}; this in turn offered more granular control over the human-agent collaborative work trajectory and prevented error propagation.

There were, however, still reasons for preferring continuous execution. P8 and P9 thought continuous execution was more efficient because it parallelized human and AI efforts: \textit{``Running all [steps] is a little bit of parallelization in my head where I can get the next thing rolling, and if I end up not having to edit it, that means we're already moving ahead''} (P8). In this case, P8 does not feel a need to be ``caught up'' with the agent's work because they know they can still steer the agent at their own pace by modifying the agent's execution and triggering re-planning at any point. 

Interleaving co-planning and co-execution via direct manipulation in particular also contributed to granular steering. Many participants, including P6, liked the precision with which they could edit the output of a particular step compared to a chat interface: \textit{``With a conversation agent, even if I'm really good at prompting it, I have to redo the prompt and keep changing that original prompt.'' (P6).} This feature also allowed P2 to quickly re-plan and repurpose the output when the agent produced unexpected (albeit helpful) paper search queries: \textit{``I was looking for more application intervention based stuff, but this is great because now I can actually use these keywords to broaden my horizon of how I was thinking about these solutions, which is I think interesting. I'm just trying to remove a couple [for future steps].''} Finally, participants found that interleaved co-planning and co-execution scoped tasks in a way that naturally afforded more precise control, especially when paired with direct manipulation, as P15 shared: \textit{``[\system] was easier to control because [the outputs] are so specific. My control is scoped down to these very small tasks. If I want to add more papers, that's easy to do.''}

\section{Deployment Study and Results}
\label{s:deployment}

In addition to interleaving \textit{co-execution and co-planning}, we were also interested in studying \textit{the delegation of human and AI work}. Thus, we tracked which steps participants assigned to the agent vs. themselves. 

In our formative study (Section \ref{s:formative-user-steps}), participants preferred to assign tasks related to information retrieval and brainstorming to the agent while keeping higher-level reasoning and synthesis tasks for themselves. In our lab study, however, most participants toggled all plan steps---even higher-level ones assigned to them by default---to the agent. 
Some avoided user steps because they were curious about ``\textit{how the bot does it}'' (P1). Understandably, testing the agent's capabilities and limitations is a natural tendency when learning to use a new system. Many speculated that they would leverage user steps if they had more time to interact with the system outside of the lab environment. P8 explicitly said that \textit{``I assign everything to [the agent] because I don't think we have that much time for me to refine my [own steps].''} 

To obtain a more holistic understanding of how \system can facilitate flexible delegation of human-AI work and more generally assist researchers day-to-day, we conducted a 7-day field deployment study with 7 participants.

\subsection{Participants and Procedure}
We invited all participants from our lab study to participate in this field deployment. 7 participants (P3, P5, P10, P11, P12, P15, P16) agreed to participate. 

Participants used \system to work on a real-world, in-progress research project of their choice over a 7-day period. Before the study, the study facilitator sent each participant a user manual\footnote{See Supplementary Materials.} documenting \system's functionalities. The manual also contained some general guidelines for the week (see Appendix \ref{a:deployment-guidelines}), including aiming to spend at least 90 minutes with the system throughout the duration of the study. Each participant could log onto their personal workspace in \system with a set of credentials provided by the study facilitator. 
We used system logs to verify that there was at least semi-regular activity for each participant throughout the study.

At the end of the 7-day period, participants completed a 30-minute semi-structured exit interview where they were asked about their general experience with the system as well as specific co-planning and co-execution interactions the study facilitator identified from within their workspace. All interviews were recorded, transcribed, and analyzed with the same qualitative analysis procedure described in Section \ref{s:data-analysis}. Participants received a \$100 USD honorarium upon completing the exit interview. 

\subsection{Field Deployment Study Results}

\subsubsection{\textbf{\system was most useful for literature synthesis and early-stage project planning}} 
\label{s:deployment-lit-review}

All participants agreed that \system was especially helpful for \textbf{literature discovery and synthesis}. This illustrates the value of interleaving human-AI efforts across planning and execution in complex information foraging and sensemaking activities. P15 found the 
structured plan for conducting literature search to be \textit{``really helpful for refining my thought process because it made me think more carefully about what I'm actually trying to look for.''} P10 agreed, and also felt that the combination of plan steps involving search query generation and paper search left them with \textit{``more faith that I was getting a good coverage of the field.''} 
P11 also appreciated \system's speed at literature search and synthesis and found it \textbf{conducive to early-stage explorations}: \textit{``if I were to come up with a summary of 20 papers, it takes probably a day, but [in \textsc{Cocoa}] it's around a minute which is really useful to formulate early-stage research ideas.''} Beyond literature search, the structure of the plan was also useful to \textit{``break down tasks and provide a sense of progress''} (P16) and \textit{``identifying important actionable steps I could take with a research problem''} (P12). Overall, participants found \system valuable in their work, with P3 in particular sharing that \textit{``I have an entire tab in my Google doc where we have stuff from \system.''} 

\subsubsection{\textbf{Step assignment strategy changed: more delegation of work with longitudinal use}}
\label{sec:field-user-steps}
Compared to the lab study, where steps were mostly assigned to the agent, participants in the field study assigned more plan steps to themselves. This was likely due to a combination of increased familiarity with the agent's capabilities and system features, as well as more opportunity to reason about which steps they wanted to keep for themselves and why. These user steps often involved high-level strategic decision-making, writing and developing arguments, running experiments and evaluations, deep reading of literature, and articulating core research ideas and novelty. 

The ability to assign steps to the agent and themselves helped participants \textbf{create a back-and-forth workflow with the agent}. For example, P15 first figured out \textit{``broad buckets of literature or themes or topics that I need to talk about''} before passing those to the agent to \textit{``help me brainstorm ways to connect those bodies of literature.''} P3, who had already conducted some interviews for their project, first had the agent look up key literature before assigning themselves a step to \textit{``pass in more of the dynamics from my interview notes''} and then had the agent propose interventions based on the literature and notes. For some participants, user steps came later in the plan because of their high-level nature. P11 often assigned the last plan step to themselves because \textit{``it was about the main innovation in the project idea [...] no one has [done it] before so I don't trust the agent.''} 

Participants also indicated that their step assignment also \textbf{depended on the stage of research they were in}. P3 admitted that \textit{``if I was just doing initial discovery and study planning, I would have gone more agent [steps].''} These findings suggest that participants benefitted from the ability to flexibly configure agency in their human-AI workflows.  

\begin{figure*}
    \centering
    \includegraphics[width=1\linewidth]{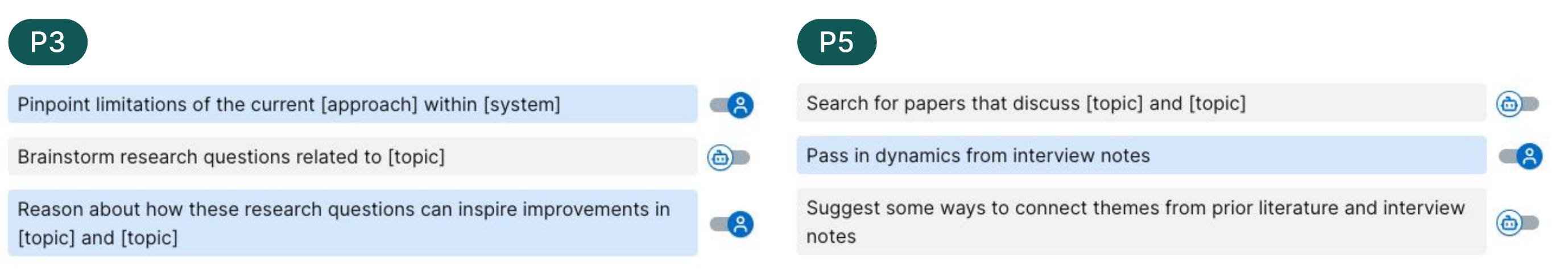}
    \caption{Examples from our field deployment where user and agent steps were interleaved in participants' plans. The plans steps were rewritten to redact project-specific details.}
    \label{fig:interleaving-field}
    \Description{Two examples of plans in Cocoa featuring alternating user- and agent-assigned steps, suggesting that user and agent efforts were interleaved.}
\end{figure*}

\subsubsection{\textbf{Interleaving co-planning and co-execution supported evolving research trajectories}}
\label{s:deployment-evolving}
Just like in the lab study, participants valued interleaving of co-planning and co-execution, but the value for supporting \textbf{flexible and iterative workflow} became more apparent with longitudinal use. As participants saw the results of execution, their understanding of the problem space shifted, which led them to want to \textbf{anchor certain steps and outputs as ``checkpoints,'' and use those as a springboard for exploring new territory}. For example, P12 shared that \textit{``I don't need to run the [earlier] retrieval step again. I know that those papers are relevant''} but instead could focus on  \textit{``only improving the second step [...] to identify something different from the literature.''} Similarly, P16 felt like they could easily ``debug'' suboptimal parts of the plan: \textit{``by forcing these checkpoints of brainstorming search queries and assembling lists of papers, it helps me get a sense of what parts of the process are good and which parts need work.''} Outputs from co-execution also inspired new approaches for participants to \textbf{``iteratively build upon a plan or reshape a plan''} (P5). Participants even mentioned that they wanted similar interleaving strategies integrated into other LLM-powered systems; for P11, it was \textit{``one of the improvements I want to see in LLM [reasoning] models''} so that users can force an early stop to the reasoning or steer it in more productive directions. Overall, our deployment of \system suggests that interleaving co-planning and co-execution helps support the non-linear nature of scientific inquiry.

\subsubsection{\textbf{Additional desiderata}}
Participants identified several desiderata for \system after spending more time with it. Some participants wanted to clearly see \textbf{what context \system used} to generate plans and execute plan steps. For example, P10 was unsure whether the system was \textit{``using the context from my other plans to generate the [other] plan.''} The system indeed was, but did not visualize relevant context in the document. P3, P15, and P16 all wanted more transparency into the shared context pool (Section \ref{s:co-execution}) used across all steps and the specific context drawn upon by each step. Several participants desired \textbf{tool interoperability}, such that they could swap out \system's default tools for their own ones (e.g., an LLM finetuned on paper summaries instead of \system's paper summarizer). This way, \system would serve as a ``hub'' for their custom tools. Finally, P5 took inspiration from Notion's ability to nest pages to suggest nested plans, where plans can be organized and expanded in a hierarchical manner. These suggestions all provide valuable directions for future work.

\section{Discussion}
\subsection{Design Considerations for Co-Agency in Human-Agent Collaboration}

Researchers have called for reframing the design of human-AI systems as not merely the specification of affordances, but construction of open design spaces that support human-AI \textit{co-agency} \cite{satyanarayan2024intelligence}. That is, rather than providing a fixed configuration of human and AI agency across all interactions, users can flexibly vary the agency they wish assert and have the AI modulate its agency in response. \system offers an early instantiation of this vision. We now reflect upon several key design decisions and discuss how future systems may build upon them or explore alternatives when designing for co-agency in human-agent systems.

We expect that \textit{fluid re-planning} will continue to be important and challenging to design for in future systems. Plans are bound to shift as the user reviews agent outputs and becomes inspired with new ideas. For example, researchers in both our lab and deployment studies dynamically re-planned with the agent upon gaining more clarity on \textit{what they actually wanted to explore} after co-executing earlier steps (see Sections \ref{s:lab-replanning} and \ref{s:deployment-evolving}). In \system, re-planning \textit{replaces} outdated steps with new ones. However, participants sometimes expressed a desire to preserve the old steps because their outputs can still be useful to reference in the future. We thus see \textit{plan branching} as a promising design pattern for exploring counterfactual plan paths. For example, systems can allow users to swap certain steps in and out of the plan, or set checkpoints at specific steps to anchor exploration. By doing so, useful work can be better retained while supporting flexible plan redirection. 

Relatedly, the \textit{length of the plan should be calibrated depending on the task}. Some tasks, such as open-ended brainstorming, benefit from shorter, partial plans where next steps can be flexibly determined based on prior execution. Indeed, a few participants expressed a preference for a step-by-step planning approach over \system's default plans when conducting brainstorming, where they did not know what next step would be until they saw the results of the previous step. On the other hand, for tasks with more structured and predictable workflows, such as literature review, participants appreciated interacting with a full plan. We calibrated the lengths of plans generated by \system based on participants' plans in our formative study. We recommend designers of human-agent systems to consider what the appropriate plan length is for tasks they hope to support, and calibrate accordingly. 

Finally, \textit{user expertise and upskilling potential should be considered} when designing for an appropriate level of agent control. As surfaced in our formative study (Section \ref{s:formative-user-steps}), there are two primary reasons why users prefer keeping a task for themselves over delegating it to the agent. \textit{First}, the user can perform the task more effectively than the agent. This is often the case where the user holds more expertise (e.g., brainstorming research questions in a user's primary research area). \textit{Second}, the task is intrinsically valuable for the user to do themselves, even if an agent is capable of automating it (e.g., reading seminal papers to develop research taste). If a human-agent system is designed for expert users who want to regularly hone their skills and intuition, providing many levels of control throughout agent planning and execution (as in \system) may be appropriate. If the system is for navigating unfamiliar domains with agents and users are not looking to upskill, fewer controls may enhance user experience. Because interactions such as co-planning and co-execution can burden the user with additional cognitive load, determining the user's capacity and willingness to take on more cognitive load can inform the appropriate level of control to afford.

These design considerations were drawn from the specific definition of agency (from Bennett et al. \cite{Bennett2023HowDH}) and domain (scientific research) specific to our study. The design space is vast, and we are excited to witness the application, refinement, and even refutation of these considerations as future work navigates this space with alternative interpretations of agency in different domains. 

\subsection{The Agent Notebook Paradigm?}

\label{s:discussion-combine}

The design of \system's interface was partly inspired by the computational notebook paradigm. Just like how notebooks help break apart a script into smaller components for unit testing and quick iteration, our ``agent notebook'' design---featuring plan step ``cells'' and options for stepwise or continuous execution---help bring similar benefits to agentic workflows. We saw in both our lab study (Section \ref{s:lab-study}) and field deployment (Section \ref{s:deployment}) that participants valued the ability to execute steps one at a time and interactively refine outputs before moving onto subsequent steps. We now discuss agent notebooks as a \textit{potentially generalizable design pattern} and consider its applications to agentic systems beyond \system.

Consider ``deep research'' systems that assemble a detailed report on a particular topic upon searching for and synthesizing information from around the web \cite{gemini-deep-research, openai-deep-research, perplexity-deep-research}. These systems may clarify user requests in a chat conversation \cite{openai-deep-research}, or allow the user to prompt the agent to regenerate the plan \cite{gemini-deep-research} before starting the research process. However, there is no room for user involvement thereafter. Agent notebooks can be particularly valuable here because they afford iteration and control over \textit{specific parts of planning and execution}. For example, the user may not be satisfied with the sources the agent is using to inform its answers and wants to steer the agent to use information from a narrow set of websites for a particular plan step. This kind of interaction is expensive without agent notebooks because it requires re-running the entire research process. Agent notebooks allow the user to focus on improving outcomes of a particular step by editing that step and rerunning it or directly editing the agent's outputs at that step, which can save significant time and cost. 

While we envision agent notebooks to be a design pattern that is generalizable beyond \system, we note that the \textit{specific affordances for co-planning and co-execution} may vary depending on the environment in which human-agent collaboration happens. \system's environment is a document editor for scientific research, and contains affordances designed around that environment (e.g., representing plan steps as editable bulleted lists, paper curation and search in the sidebar). Translating these affordances to coding, for example, may need to draw upon different interaction metaphors. The plan can be represented as an executable spec file (e.g., a special Jupyter notebook with only markdown cell blocks that serve as step instructions), while co-execution can resemble live pair programming. We anticipate that making intentional design choices for these affordances by drawing from appropriate metaphors will be an important part of future work.

\subsection{The Practical Significance of User Steps and Planning Ahead}

In \system, users are presented with an interactive multi-step plan up front, which the agent adheres to and can modify while operating. Importantly, this plan also allows users to flexibly delegate a step to themselves or the agent. Both design decisions are a departure from many existing agentic AI systems, where the agent dynamically constructs a plan step-by-step as it generates outputs \cite{Yao2022ReActSR} and seeks user input only when it \textit{automatically} determines it should do so \cite{openai-operator, Peng2025MoraePP, Shao2024CollaborativeGA}. We discuss two practical reasons---cost and oversight---for why such a departure may be desirable for practical agents in the real world.

AI agents can be prohibitively expensive to run \cite{Kapoor2024AIAT, Yang2024SWEagentAI}. Because an agent's actions often involve multiple LLM calls, costs can quickly accumulate, especially for more complex and long-running tasks. In response, academic projects have implemented cost-capping measures to alleviate this \cite{Yang2024SWEagentAI}, and researchers have called for agent evaluations to be cost-controlled \cite{Kapoor2024AIAT}. In user-facing interactive systems, cost also includes time spent waiting for agents to come back with results. These cost considerations change user behavior. When participants made the decision to assign most or all steps to the agent in our lab studies, they viewed the agent's actions as rather inconsequential and low-cost---they could run the agent and see what happens, and edit the output post hoc (Section \ref{s:lab-replanning}). However, while this was the case in the lab, it may not be true in some real-world contexts where users and/or developers need to pay monetary and time costs for each agent step. User steps become an important cost-saving measure. Users can choose to complete a step that is not particularly burdensome or is too difficult for current AI agents to complete effectively. Efficiently leveraging human effort and expertise where appropriate via user steps can be a key step towards developing cost-aware agents that are usable and affordable in practice.

The increased autonomy of AI agents also comes with increased risks, as AI harms become more difficult to anticipate, and accountability for AI actions becomes harder to trace \cite{chan2023harms, chan2024visibility}. \system's interface not only provides more transparency into agent actions, but also provide some assurance about and control over the agent's execution trajectory. By generating a plan ahead of execution that the agent adheres to, harm anticipation and mitigation become more tractable. Step-by-step plan generation used by existing agent frameworks, such as ReAct \cite{Yao2022ReActSR}, may also accomplish a similar goal by asking for user approval whenever a new step is generated. However, this can severely limit agent autonomy when desirable and may also disengage the user if their only form of supervision is repetitive rubber-stamping \cite{Kapoor2024AIAT}. 
Additionally, advanced planning enables assessments of task risk pre-execution. Plan steps determined to be of higher risk (e.g., requiring working with passwords) can be assigned to the user---even if the agent is capable of completing them---as a risk mitigation measure.

\section{Limitations and Future Work}
\label{s:limitations-fw}
Based on our formative study, we built \system as a document editor because documents are natural sites of planning for researchers and present an ideal environment for agent interaction (Section \ref{s:formative-documents}). However, the linear nature of a document also has its limitations and does not support  ``forking'' plans and iterating on versions of a plan in parallel. These types of interactions are better supported by node-based canvases, which have been gaining popularity as a means of sensemaking and creative exploration with LLMs (e.g., \cite{Suh2023SensecapeEM, Pu2024IdeaSynthIR, Arawjo2023ChainForgeAV}). Future work can explore new interaction techniques for interleaved co-planning and co-execution. The default assignment of user and agent steps in \system is also a result of researchers' preferences from our formative study. Past work explored automated methods for a model to decide when to defer a task to a human expert versus acting on its own \cite{mozannar2020consistent}. Future work may investigate applications of these methods in human-agent systems and/or leverage inference-time scaling \cite{openai-o1, muennighoff2025s1} to improve plan quality. 

Our baseline mimicked the dominant interaction paradigm in AI-powered research tools: chatbots. While this allowed us to elicit participant feedback on the usability of \system compared to the status quo, we could have designed additional baselines by ablating parts of our design. For example, our baseline could have been an interactive planning interface that did not support stepwise execution. In future work, we  can expand our participant pool and run more studies with ablations of \system to pinpoint which particular affordances participants valued most. Further, the underlying agent used in \system also had some technical limitations. Since it was not multimodal, it was not capable of taking visual content (e.g., figures or slides) as input. The agent was also not capable of executing code---a capability which a couple of participants inquired about during the study. 
Future work can expand \system's utility with multimodal agents and/or agents that can execute code. 

Another limitation is the demographics of our participants. Our participants were researchers in CS and CS-adjacent areas. Research culture and incentives specific to CS may bias our results and limit our imagination of how \system can be used. P10, who used to work in wet labs in their undergraduate research, shared that a promising use case for \system would be helping wet lab researchers walk through the steps necessary to prepare for lab experiments. Future work may investigate novel applications of systems like \system in domains beyond CS.

\section{Conclusion}

In this paper, we presented \system, an interactive system for scientific researchers to flexibly collaborate with an AI agent in a document editing environment to tackle open questions and tasks within their research projects. Motivated by a formative study with 9 researchers, \system introduced a new approach for human-agent collaboration---\emph{interleaved co-planning and co-execution}---by which a human and agent can iteratively create plans of action and execute the plan in a mixed-initiative manner that  centers human agency. 

Our lab ($n=16$) and field deployment ($n=7$) studies showed that participants benefited from interleaved co-planning and co-execution in the lab and in the field. Compared to a strong baseline with a more familiar chat interface, \system enhanced agent steerability without sacrificing ease of use. In the lab study, participants' main mode of co-execution was refining agent outputs, but when \system was integrated into their research workflows in the field over a 7-day period, participants took advantage of user steps to guide the agent with their expertise and exert more of their agency. Overall, our work demonstrates the promise of our approach for flexibly mediating human and AI agency. Importantly, our work offers many practical takeaways for the design of human-agent systems, such as leveraging user steps and planning ahead for agent cost management and oversight.

AI agents have exciting potential to transform our digital experiences and advance long-standing visions held by HCI and AI communities. However, these transformations and advancements also demand renewed attention to strategies for ensuring effective and flexible human-agent collaboration. Our work advances efforts in this emerging frontier.

\begin{acks}
We warmly thank all our participants from our formative, pilot, lab, and deployment studies. We thank our anonymous reviewers for feedback on our manuscript. Finally, we thank Raymond Fok and Shannon Shen for helpful discussions.
\end{acks}

\bibliographystyle{ACM-Reference-Format}
\bibliography{refs}


\newpage
\appendix

\section{Formative Study Participants}
\label{a:formative-study-participants}
See Table \ref{t:participants-formative}

\begin{table*}[h!]
\small
\centering
    \begin{tabular}[h]{p{1cm} p{1cm} p{1.5cm} p{1.5cm} p{1.8cm} p{1.8cm} p{1.7cm} p{2cm}}
    \toprule
    P\# & Gender & Age Range & Country & Research YoE & Research Area & Research Area (General) & AI Use Frequency\\
    \midrule 
    P1 & Man & 25--34 & U.S. & 6--10 & Human-AI interaction & HCI & A couple times\\
    P2 & Man & 25--34 & U.S. & 6--10 & LLM evaluation & NLP & Occasionally\\
    P3 & Woman & 18--24 & U.S. & 2--5 & LLMs + society & NLP & Never\\
    P4 & Man & 25--34 & Canada & 2--5 & Multilingual NLP & NLP & Occasionally\\
    P5 & Woman & 25--34 & U.S. & 6--10 & Human-AI interaction & HCI & Occasionally\\
    P6 & Man & 25--34 & U.S. & 2--5 & Multimodal AI \& HCI & HCI & Occasionally\\
    P7 & Woman & 25--34 & South Korea & 2--5 & LLM retrieval & NLP & Occasionally\\
    P8 & Man & 25--34 & U.S. & 6--10 & Human-AI interaction & NLP & Occasionally\\
    P9 & Woman & 18--24 & U.S. & 0--1 & Creativity support tools & HCI & A couple times\\
    \bottomrule
    \end{tabular}
    \caption{Participants from our formative study. All participants were Ph.D. students. \textbf{Country} refers to the country in which the participant primarily conducts research at the time of the study. \textbf{Research YoE} refers to the years of experience conducting academic research. \textbf{AI Use Frequency} refers to how frequently the participants uses AI tools to ideate and/or iterate on research ideas. Participants selected ``Occasionally'' based on the description ``I don't rely on AI but sometimes tinker with it.'' }
    \label{t:participants-formative}
    \Description{Table with the participant ID, gender, age range, country in which they conduct their research, years of experience conducting research, specific research area, general research area, and AI use frequency for each participant from our formative study.}
\end{table*}

\section{\system Formative Study Procedure Details}
\label{a:formative-procedure-details}
We provide more details of each portion of our formative study and its associated activities below.

\subsection{Project document activity}

First, we asked participants to discuss some of the ideas in their project document they shared with us, the origins of those ideas, and the ideal ways they envision AI-powered tools helping them throughout the research process. We then invited participants to a Google Doc or Slides that we converted from their project document for easy collaboration; this conversion was made by copying and pasting content from their project document if it was not already in Docs or Slides. We asked participants to brainstorm 2--3 remaining questions they have about their project that they would be interested in tackling. 
For each question, participants created a bulleted plan they would undertake to pursue it, which we then used to answer \textbf{F2}.

\subsection{Probe activity}
\label{s:formative-probe-activity}
In the second activity, we presented participants with a WoZ design probe in the form of two side-by-side Google Docs. One Doc, the ``planning document,'' contained 3 research ideas generated by Perplexity AI \cite{perplexity} based on the description of research interests submitted by the participant and 1--3 of their most recent publications and/or preprints. Participants were asked to work in the doc to expand upon an idea (or combine multiple) through a self-defined plan, with the option of invoking the help of an AI assistant through requests prefaced with a ``!'' command. The study facilitator (the Wizard) entered this command into Perplexity AI and pasted the output to the other Doc, the ``AI output document.'' While waiting for the response, we asked participants to write a brief plan consisting of 3--6 steps for how they would complete the request themselves. Participants then inspected the assistant's output and incorporated any useful text into their document. This process repeated until the participant felt like their idea was sufficiently concrete to write a short descriptive paragraph about it, typically after at least 2--3 requests to the AI. This activity helped us answer \textbf{F3}.

\subsection{Concluding interview}
The study ended with a concluding interview, where we asked participants about their experiences, perceptions, and desiderata for the AI assistant, along with preferred ways to interact with it. Each participant was given a \$35 USD honorarium after the study. The study was reviewed and approved by the Allen Institute of AI's IRB. 

\section{Formative Study Data Analysis}
\label{a:formative-data-analysis}
To answer \textbf{F1}, the first author used a hybrid inductive-deductive coding process \cite{fereday2006demonstrating} to code participants' submitted project documents. The first 5 documents were inductively coded to surface common structural elements before the elements were deductively applied to the remaining 4. We iterated on existing elements and added new ones as needed. 

For \textbf{F2}, we performed inductive thematic analysis on two documents---participants' submitted project documents and in-study planning documents. The first author sourced \textit{intentions for planning} within submitted project documents by locating the snippet of text that initiates a plan. For example, if the plan consists of a bulleted list, the intention is often expressed in the lead-in text that immediately precedes the list. We note that an individual plan item can also signal planning intent if it contains a nested plan. The first author performed thematic analysis by inductively coding the extracted text. The first author then identified and extracted \textit{plan steps} participants wrote in both documents before inductively coding them. 

For \textbf{F3}, the first author performed open coding on the study transcripts before thematically analyzing the coded snippets. To enrich the data, the first author also extracted and inductively coded all requests to the assistant from participants' planning documents. The codes were discussed and iterated upon with other project team members on a weekly basis over the course of 3 weeks. 

\section{Further Implementation Details}
\label{a:implementation}

\subsection{Underlying LLM Agent}

\subsubsection{Research support tools}

We created a custom tool-calling LLM agent\footnote{\url{https://platform.openai.com/docs/guides/function-calling}} powered by GPT-4o.
While GPT-4o provides APIs for managing multi-turn conversation context, the system manages its own context using a JSON schema we developed for storing key insights from papers and other forms of output without storing the full artifact. Previous plan steps and their outputs are also represented in the schema to enrich context.  
To provide high-quality research support, we incorporated two best-in-class research support functionalities into our agent. The first tool allowed the agent to access The Semantic Scholar Open Data Platform \cite{Kinney2023TheSS} via its publically available search APIs.\footnote{https://www.semanticscholar.org/product/api} This tool provided the agent access to the full text and metadata (such as authors and publication venues) of millions of research papers, created by a state-of-the-art PDF extraction and knowledge graph normalization pipeline, allowing the agent to search for authors and papers based on topics, titles, and rich metadata \cite{Kinney2023TheSS}. The second tool is the \emph{Ask this Paper}\footnote{https://www.semanticscholar.org/product} feature on Semantic Scholar which generates paper summaries based on the paper's full text given an aspect of interest.
This tool allowed the agent to summarize a relevant paper based on specific aspects of interest within a plan, and also synthesize specific aspects across many papers.
For example, the agent can use the first tool to find papers relevant to a topic of interests from a particular author, then, used the second tool to summarized the findings of each papers, and, finally, synthesize the individual summaries into a compare and contrast statement about the set of papers for the user to review.  All prompts used are provided in our Supplementary Materials.

\subsection{Guidance for plan generation}
We use findings from our formative study to guide the agent's generation of initial plans. In our formative study, we had asked participants to write down brief plans when interacting with our probe (see Appendix \ref{a:formative-procedure-details}). 
Based on participants' preferences of which steps they prefer assigning to an AI vs. keeping for themselves (\textbf{F3} from our formative study), we organized these steps into agent and user steps. We then wrote examples of how these steps are composed into plans for tackling particular questions, once again drawing from plans participants wrote in the probe activity. These examples were provided for in-context learning in \system's system prompt for plan generation (see prompts in our Supplementary Materials).
Finally, we also enable \system to learn from interactive plans the user has previously created and edited. When the agent is invoked, \system collects the existing plans in the document and adds them to the in-context learning examples on-the-fly. This allows the user's co-planning efforts to be reused for similar requests elsewhere in the document. 

We also enable the agent to learn from previously created plans by extracting existing plans in the document and integrating them into system context. 

\section{Lab Study Participants}
\label{a:user-study-participants}
See Table \ref{t:participants-eval}

\begin{table*}[h!]
\small
\centering
    \begin{tabular}[h]{p{0.3cm} p{1cm} p{1.2cm} p{1.2cm} p{2cm} p{1cm} p{2.8cm} p{1.7cm} p{2cm}}
    \toprule
    P\# & Gender & Age Range & Country & Job Title & Research YoE & Research Area & Research Area (General) & AI Use Frequency\\
    \midrule 
    P1 & Woman & 25--34 & U.S. & Ph.D. student & 2--5 & Social computing & HCI & Frequently\\
    P2 & Woman & 25--34 & U.S. & Ph.D. student & 2--5 & Social computing & HCI & Occasionally\\
    P3 & Woman & 25--34 & U.S. & Ph.D. student & 6--10 & Visualization & HCI & Occasionally\\
    P4 & Woman & 25--34 & U.S. & Ph.D. student & 2--5 & Culture + computing & HCI & A couple times\\
    P5 & Woman & 25--34 & U.S. & Postdoc & 6--10 & Software development & HCI & Occasionally\\
    P6 & Woman & 25--34 & U.S. & Ph.D. student & 6--10 & Health + computing & HCI & Occasionally\\
    P7 & Man & 25--34 & U.S. & Ph.D. student & 2--5 & LLMs & NLP & Frequently\\
    P8 & Woman & 25--34 & U.S. & Ph.D. student & 2--5 & Accessibility & HCI & Never\\
    P9 & Man & 25--34 & U.S. & Ph.D. student & 2--5 & Multimodal AI & ML & Frequently\\
    P10 & Woman & 18--24 & U.S. & Ph.D. student & 2--5 & On-device AI & ML & A couple times\\
    P11 & Man & 18--24 & U.S. & Ph.D. student & 2--5 & Ubiquitous computing & HCI & Always\\
    P12 & Man & 25--34 & U.S. & Ph.D. student & 6--10 & LLM evaluation & NLP & Frequently\\
    P13 & Man & 35--44 & Canada & Ph.D. student & 2--5 & Computational biology & ML & Frequently\\
    P14 & Woman & 18--24 & U.S. & Ph.D. student & 2--5 & LLMs & NLP & Frequently\\
    P15 & Woman & 25--34 & U.S. & Ph.D. student & 2--5 & Social computing & HCI & Occasionally\\
    P16 & Man & 25--34 & U.S. & Postdoc & 6--10 & Human-AI interaction & HCI & Occasionally\\
    \bottomrule
    \end{tabular}
    \caption{Participants from our user study. \textbf{Country} refers to the country in which the participant primarily conducts research at the time of the study. \textbf{Research YoE} refers to the years of experience conducting academic research. \textbf{AI Use Frequency} refers to how frequently the participants uses AI tools to ideate and/or iterate on research ideas. Participants selected ``Occasionally'' based on the description ``I don't rely on AI but sometimes tinker with it.'' }
    \label{t:participants-eval}
    \Description{Table with the participant ID, gender, age range, country in which they conduct their research, job title, years of experience conducting research, specific research area, general research area, and AI use frequency for each participant from our lab study.}
\end{table*}

\section{Lab Study Recruitment Process and Pilot Studies}
\label{a:user-study-recruitment}
We sent out a recruitment form to collect basic demographic details, frequency of AI use in research, and a copy of a project document from an ongoing or completed project document to use during the study. One participant also participated in our formative study. 15 researchers were based in the United States, and 1 was based in Canada. Participants' research areas ranged from large language models, to ubiquitous computing, to visualization; broadly, they spanned HCI ($n=10$), ML ($n=3$), and NLP ($n=3$). In our participant pool, many users were occasional (6) or frequent\footnote{We define ``occasional'' and ``frequent'' the same way as we do in our formative study (Table \ref{t:participants-formative}).} (6) users of AI tools in the research process. 

We also recruited 4 additional participants (1 female, 2 male, 1 non-binary) from our institution for pilot studies to test out \system, catch usability issues, and help us refine our procedure. We learned from our pilot studies that the length of generated plans should be around 3 steps (rather than the system's default of 5--6 steps) to fit within the study's time constraints. 

\section{Lab Study Procedural Details}
\label{a:user-study-details}
To control for the length of the study, the first author reviewed the participants' documents prior to the study to identify and highlight two candidate tasks. To make sure the two tasks are valid and comparable, early on in the study, the first author asked participants 1) whether the tasks have been decisively solved by the participant and whether they would still like to explore them in the study, and 2) whether the two tasks are similar in scope, such that one task does not take significantly more resources and effort to explore than the other. If these conditions were not met, the participant was prompted to rewrite one or both tasks until they were. The two tasks were then randomly assigned to \system and the baseline. 

In the \system condition, participants first watched a 6-minute video demo of the system, and were then directed to a document where they would get used to using \system's features on the following sample question: \textit{``How can we use AI to better society?''} This practice period lasted around 10 minutes. Participants were then asked to spend 25 minutes to make as much progress as possible tackling the task assigned to this condition with \system, thinking aloud as they did so. They were also permitted to also use other tools (academic search engines, other AI tools, social media, etc.) alongside \system, although very few did. After the 25 minute session, participants wrote a brief set of takeaways and next steps from their exploration, and filled out a short evaluation form with 5-point Likert scale questions about their experience (see Appendix \ref{a:eval-likert-form} for this form). After submitting the form, they were encouraged to further try out \system on other parts of their project document for another 10--15 minutes. 

In the baseline condition, we did not provide participants a tutorial because the chat interface was already ubiquitous beyond interacting with AI agents. Participants were once again asked to spend 25 minutes making as much progress as possible tackling the task assigned to this condition, thinking aloud as they did so and using other tools if desired. They wrote brief takeaways and next steps after the 25 minutes was up and filled out the same evaluation form as the \system condition.

Because participants identified literature search and understanding as one of the areas they can benefit most from AI assistance (Section \ref{s:formative-planning}), we had participants tackle a literature-augmented task in both \system and the baseline. By \textit{literature-augmented}, we mean tasks that are to be completed by referencing or drawing from academic literature; in the study, participants worked on not only literature review tasks, but also tasks where researchers needed to make literature-informed decisions such as study design and ideation. This stands in contrast with tasks focused on writing mechanics (e.g., rewording a sentence) or tasks that do not involve literature (e.g., booking travel) that past work has covered (e.g., \cite{yao2022webshop, Yang2024SWEagentAI, kim2023metaphorian, zhang2023visar, Ma2023BeyondCE, Lee2024ADS}). 

\section{Lab Study Self-Evaluation Form Questions}
\label{a:eval-likert-form}
All questions were answered on a 5-point Likert scale.
\begin{itemize}
    \item The system’s outputs were useful to me for exploring the research problem at hand
    \item I obtained new and useful insights from using the system
    \item The system helped me clearly understand the steps needed to effectively tackle this particular problem
    \item By using the system, I’ve developed a better understanding of potential solutions to this particular problem
    \item I found the system easy to use
    \item I could easily steer the system towards doing something helpful \looseness=-1
    \item I could see myself easily integrating this system into my workflow \looseness=-1
    \item I’m satisfied with the artifact (summary and next steps) that I’ve created after using the system
    \item I feel confident about my next steps after using the system
\end{itemize}

\section{Deployment Study General Guidelines}
\label{a:deployment-guidelines}
\begin{enumerate}
    \item Try to spend a total of at least 90 minutes with Cocoa during the 7-day use period.
    \item It may help to jot down any notable experiences with Cocoa so we can more easily discuss them later.
    \item Treat the document as a scratch space to tinker with and explore rough ideas.
    \item Cocoa is just another tool in your toolbox. You can use it alongside whatever you typically use in your workflows!
    \item If the system crashes at any point, try refreshing the page. The document auto-saves regularly.
\end{enumerate}


\end{document}